\documentstyle[12pt,a4,psfig,epsf]{article}
\begin{document}
\renewcommand {\thepage} { }
\renewcommand {\thefootnote} {\fnsymbol{footnote}}
\setcounter {page} {0}
\setcounter {footnote} {0}
\vspace*{1cm}
\vspace{5mm}
\begin{center}{\Large \bf
Quantum optical input--output relations for dispersive and lossy
multilayer dielectric plates
}\end{center}
\vspace{5mm}
\begin{center}
T. Gruner and D.-G. Welsch
\\
Friedrich-Schiller Universit\"at Jena,
Theoretisch-Physikalisches Institut\\ Max-Wien Platz 1,
D-07743 Jena, Germany
\end{center}
\vspace{3cm}
\begin{center}{\bf Abstract}\end{center}
Using the Green function approach to the problem of
quantization of the phenomenological Maxwell theory,
the propagation of quantized radiation through dispersive and absorptive
multilayer dielectric plates is studied. Input--output relations are
derived, with special emphasis on the determination of the
quantum noise generators associated with the absorption of
radiation inside the dielectric matter. The input--output
relations are used to express arbitrary correlation functions of the
outgoing field in terms of correlation functions of the incoming field
and those of the noise generators. To illustrate the theory, the effect
of a single-slab plate on the mean photon-number densities in the
frequency domain is discussed in more detail.
\vfill\newpage
\renewcommand {\thepage} {\arabic{page}}
\setcounter {page} {1}
\section{Introduction}
\label{intro}
It is well known that the use of instruments in optical experiments
needs careful examination with regard to their action on the quantum
statistics of the light under study. In particular the presence of
passive instruments that may be regarded as macroscopic dielectric
bodies responding linearly to radiation can be included in the theory
by quantization of the phenomenological Maxwell theory for radiation in
linear inhomogeneous dielectrics. The formalism was first developed for
dispersionless and lossless dielectrics \cite{1,2,3,3a} and successfully
applied to the study of the action of various devices, such as dielectric
plates and interfaces \cite{4,5} and optical cavities \cite{6,7}. When
broad-band radiation, such as an ultrashort light pulse, propagates
through (long-distance) dielectric devices the effects of dispersion
and absorption, which are related to each other by the
Kramers--Kronig relations, must necessarily be taken into account.
In this context, a number of questions have been arisen,
such as the question about the velocity at which a single-photon
wavepacket travels in dielectric matter under the influence of
normal and/or abnormal dispersion in the vicinity of an absorption
line \cite{7a,7b}. Needless to say that absorption also introduces
additional noise, at least the (quantum) vacuum noise.

The problem of describing the effects of dispersive and absorptive
linear dielectric devices on quantized radiation has been considered in a
number of papers and various quantization schemes have been
proposed \cite{8,8a,9,10,10a,11,12,13}. As we have recently shown,
quantization of the radiation field within the frame of
the phenomenological Maxwell theory (with given complex permittivity
in the frequency domain) can be performed using a Green-function
expansion of the operator of the vector potential \cite{12,14}.
This quantum-field-theoretical formalism, which may be regarded as a
generalization of the familiar concepts of mode expansion, applies to both
homogeneous and inhomogeneous dielectric matter and is consistent
with both the Kramers--Kronig relations and the canonical
(equal-time) field commutation relations in QED.

In the present paper we use the method in order to study the
behaviour of quantized radiation in the presence of a multilayer
dielectric plate and to derive input--output relations.
The calculation of input--output relations is commonly
based on a development of the familiar formalism of quantum
noise theory (see, e.g. \cite{14a}), in which the explicit nature
of the input from a heat bath, and the output into it, is taken
into account. The advantage of the quantum-field-theoretical approach
is that it enables one to obtain the input--output relations
including their dependence on frequency through geometry and
dispersive and absorptive properties of the layers, because
it consistently accounts for the effects of radiation-field
propagation according to the (phenomenological) quantum Maxwell
equations.

It is well known that when dielectric matter in free space is considered
and the imaginary part of the permittivity may be disregarded (provided
that the losses in the chosen frequency interval are sufficiently small),
then the input--output relations correspond to unitary transformations
between input- and output-mode operators \cite{15,16}.
These concepts fail when the effect of absorption of radiation
through the dielectric matter is taken into account. In this
case the outgoing fields are not only related to the incoming
fields but also to appropriately chosen noise sources \cite{17}.
Neglection of these supplementary contributions would unavoidably
lead to a violation of the canonical commutation relations for
the outgoing fields, and hence effects of quantum noise
would left out. Knowing the input--output-relations, the quantum
statistics of the outgoing fields can be obtained from the statistics
of the incoming fields and internal (temperature-dependent) noise
properties of the dielectric matter under study.

The input--output relations derived may be used in various
quantum-optical applications, because the multilayer dielectric
structure under study may serve as a model for a number of devices,
such as mirrors, beam splitters, interferometers of Fabry-P\'{e}rot
type, and optical fibres. In particular, multilayer dielectric
mirrors may be regarded as tunnel barriers for photons \cite{17a} and
can be used in order to measure barrier traversal times \cite{7a}.
The apparently superluminal behaviour of photons at dielectric
tunnel barriers, which has been observed in recent experiments,
may be modified by the absorption in the dielectric structures \cite{17b}.
These and related studies have been a subject of increasing
interest in order to answer a fundamental question of quantum
physics, namely, of how much time it takes a particle to tunnel
through a barrier.

The paper is organized as follows. In Sec.~\ref{sec2} the
Green function approach to the quantization of radiation
in dispersive and lossy linear dielectrics is summarized.
Applying this quantization scheme to radiation propagating
through multilayer dielectric plates, in Sec.~\ref{inout}
input-output relations are derived. In Sec.~\ref{corrfunc}
these relations are used to study normally ordered correlation
functions of the outgoing fields. To illustrate the theory,
results for the photon-number density statistics observed
in the case of a single-slab dielectric plate are given.
Finally, a summary and some conclusions are given in Sec.~\ref{sum}.

\section{Quantization scheme}
\label{sec2}
\subsection{Green's function approach}
\label{sec2.1}
Let us consider linearly polarized radiation propagating in
$x$\,direction and allow for the the presence of a multilayer dielectric
plate characterized in terms of a frequency-dependent permittivity
$\epsilon(x,\omega)$ that varies in $x$\,direction and
obeys, for causality reasons, the Kramers--Kronig relations.
Introducing the vector potential $A(x,t)$ and using the relation
\begin{equation}
D(x,t) = \epsilon_0 \left[
E(x,t) + \int_{-\infty}^{t} {\rm d} \tau \, \chi (t-\tau) E(x,\tau)
\right]
\label{2.2}
\end{equation}
($E\!=\!-\dot{A}$), the classical phenomenological Maxwell
equations yield
\begin{equation}
\frac{\partial^2}{\partial x^2} A (x,t) - \frac{1}{c^2} \! \left[
\ddot{A}(x,t) + \! \int_{-\infty}^t \! {\rm d} \tau \,
\chi(t-\tau) \ddot{A} (x,\tau)
\right] \! = 0,
\label{2.3}
\end{equation}
which in the frequency domain reads as
\begin{equation}
\left[ \frac{\partial^2}{\partial x^2}
+ \frac{\omega^2}{c^2} \epsilon(x,\omega) \right] A(x,\omega) = 0.
\label{2.6}
\end{equation}

Equation~(\ref{2.6}) may be transferred to quantum theory
as follows \cite{12,14}:
\begin{equation}
\left[ \frac{\partial^2}{\partial x^2}
  + \frac{\omega^2}{c^2} \epsilon(x,\omega) \right] \hat{A}(x,\omega) =
\hat{\!j}(x,\omega),
\label{2.6a}
\end{equation}
where the ``current'' operator
\begin{equation}
\hat{\!j}(x,\omega) = \frac{\omega}{c^2}
\sqrt{\frac{\hbar}{\pi \epsilon_0 {\cal A}}\,\epsilon_{\rm i}(x,\omega)}
\,\hat{f}(x,\omega)
\label{2.14}
\end{equation}
is introduced to take into account the noise owing to absorption.
In Eq.~(\ref{2.14}), $\hat{f}(x,\omega)$ and $\hat{f}^\dagger(x,\omega)$
are bosonic field operators,
\begin{equation}
\left[
\hat{f}(x,\omega) , \hat{f}^{\dagger} (x',\omega')
\right] = \delta(x-x') \, \delta(\omega - \omega'),
\label{2.16}
\end{equation}
\begin{equation}
\left[
\hat{f}(x,\omega) , \hat{f} (x',\omega')
\right] = \left[
\hat{f}^{\dagger}(x,\omega) , \hat{f}^{\dagger} (x',\omega')
\right] = 0,
\label{2.17}
\end{equation}
and
\begin{equation}
\epsilon(x,\omega) = \epsilon_{\rm r}(x,\omega)
                        + i \, \epsilon_{\rm i}(x,\omega)
\label{2.15}
\end{equation}
(${\cal A}$, normalization area in the $yz$\,plane).
The solution of Eq.~(\ref{2.6a})
may be given by
\begin{equation}
\hat{A}(x,\omega) = \int {\rm d} x'\, G(x,x',\omega)
\,\hat{\!j}(x',\omega),
\label{2.8}
\end{equation}
where $ G(x,x',\omega) $ is the classical Green function obeying
the equation
\begin{equation}
\left[ \frac{\partial^2}{\partial x^2}
+ \frac{\omega^2}{c^2} \epsilon(x,\omega) \right] G (x,x',\omega) =
\delta(x-x')
\label{2.9}
\end{equation}
and tending to zero as $x\!\to\!\pm\infty$.
It can be shown \cite{12,14}, that the quantization scheme outlined
above ensures that the (Schr\"{o}dinger) operators of the vector
potential and the electric-field strength,
\begin{eqnarray}
\lefteqn{
\hat{A}(x) = \int_0^{\infty} {\rm d} \omega\, \hat{A}(x,\omega) +
{\rm H.c.},
}
\label{2.11}
\\ & &
\hat{E}(x) = i \int_0^{\infty} {\rm d} \omega\, \omega
\hat{A}(x,\omega)
+ {\rm H.c.},
\label{2.12}
\end{eqnarray}
satisfy the well-known canonical commutation relation
\begin{equation}
\left[
\hat{A}(x),\hat{E}(x')
\right] = - \frac{i \hbar}{{\cal A} \epsilon_0} \delta(x-x').
\label{2.10}
\end{equation}

\subsection{Quantum Langevin equations}
\label{sec2.2}
 From Eqs.~(\ref{2.11}), (\ref{2.8}), and (\ref{2.9}) we see
that the problem of determining the operator of the
vector potential reduces to the calculation of the
classical Green function. Before performing the calculations
for multilayer dielectric plates, let us summarize some results
for homogeneous dielectrics, where
\begin{eqnarray}
G(x,x',\omega) & = &
\frac{1}{2\pi} \int_{-\infty}^\infty {d}k\, e^{ik(x-x')}
\frac{c^2}{\omega^2\epsilon(\omega)-c^2k^2}
\nonumber \\
& = & \left[2i\frac{\omega}{c}n ( \omega )\right]^{-1}
\exp\!\left[ i \frac{\omega}{c} n ( \omega )\,|x-x'| \right]\!.
\label{2.10a}
\end{eqnarray}
Here
\begin{equation}
n(\omega) = \sqrt{\epsilon(\omega)}
= \beta(\omega) + i \, \gamma(\omega)
\label{2.10b}
\end{equation}
is the complex refractive index. Using Eqs.~(\ref{2.8}) and (\ref{2.10a}),
Eq.~(\ref{2.11}) may be rewritten as
\begin{equation}
\hat{A}(x) = \int_0^{\infty} {\rm d} \omega\,
\sqrt{\frac{\hbar}{4 \pi c \omega \epsilon_0 \beta (\omega) {\cal A}}}
\,\frac{\beta(\omega)} {n(\omega)}
\left[
e^{i \beta(\omega) \omega x/c} \hat{a}_+ (x,\omega)
+ e^{-i \beta(\omega) \omega x/c} \hat{a}_- (x,\omega)
\right] \!+\! {\rm H.c.},
\label{2.18}
\end{equation}
where the operators
\begin{equation}
\hat{a}_{\pm} (x,\omega) = \frac{1}{i} \sqrt{2 \gamma(\omega)
\frac{\omega}{c}} \,
e^{\mp \gamma(\omega) \omega x/c}
\int_{-\infty}^{\pm x} {\rm d} x'\,
e^{-i n(\omega) \omega x'/c} \hat{f} (\pm x', \omega)
\label{2.20}
\end{equation}
associated with the waves propagating to the right ($+$) and left ($-$)
are introduced. In the limiting case when the absorption may be disregarded
[$\gamma(\omega)\omega|x\!-\!x'|/c \!\to\!0$], Eq.~(\ref{2.18})
reduces to the familiar mode-expansion result \cite{12,14}.
In particular, the operators $\hat{a}_\pm(x,\omega)$ and
$\hat{a}_\pm^\dagger(x,\omega)$ become independent on $x$ and satisfy
the well-known commutation relations for photon destruction and
creation operators, respectively.

Equation~(\ref{2.20}) implies that the operators $\hat{a}_\pm(x,\omega)$
and $\hat{a}_\pm(x',\omega)$ can be related to each other as
\begin{equation}
\hat{a}_{\pm} (x,\omega) = \hat{a}_{\pm} (x',\omega)
e^{\mp \gamma (\omega) \omega (x - x') / c}
+ \int_{x'}^x {\rm d} y\, \hat{F}_{\pm} (y, \omega)
e^{\mp \gamma (\omega) \omega (x - y)/c},
\label{2.27}
\end{equation}
where
\begin{equation}
\hat{F}_{\pm} (x, \omega) = \pm \frac{1}{i} \sqrt{2 \gamma(\omega)
\frac{\omega}{c}} \,
e^{\mp i \beta(\omega) \omega x/c}
\hat{f} (x, \omega).
\label{2.24}
\end{equation}
Hence, the operators $\hat{a}_\pm(x,\omega)$ obey quantum
Lange\-vin equations in the space domain \cite{14},
\begin{equation}
\frac{\partial}{\partial x} \hat{a}_{\pm} (x, \omega)
= \mp \gamma(\omega) \frac{\omega}{c} \hat{a}_{\pm} (x, \omega)
+ \hat{F}_{\pm} (x, \omega),
\label{2.23}
\end{equation}
where the operators $\hat{F}_\pm(x,\omega)$ may be regarded as
Lan\-ge\-vin noise sources,
\begin{equation}
\left[
\hat{F}_{\pm} (x, \omega) , \hat{F}_{\pm}^{\dagger} (x', \omega')
\right] = 2 \gamma (\omega) \frac{\omega}{c} \delta(x \!-\! x')
\delta(\omega \!-\! \omega'),
\label{2.25}
\end{equation}
\begin{equation}
\left[
\hat{F}_{\pm} (x, \omega) , \hat{a}_{\pm}^{\dagger} (x', \omega')
\right]
= \pm \Theta \left( \pm x' \mp x \right)
 2 \gamma (\omega) \frac{\omega}{c}
\, e^{- \gamma(\omega) \omega| x' - x |/c}
\delta(\omega - \omega').
\label{2.26}
\end{equation}

\section{Input--output relations}
\label{inout}
We now turn to the problem of propagation of quantized radiation through
multilayer dielectric plates. Assuming $N$ dielectric slabs, the interfaces
being parallel to the $yz$\,plane
(cf. Fig.~1),
the permittivity may be given by
\begin{equation}
\epsilon(x,\omega) = \sum_{j=1}^N \lambda_j(x) \epsilon_j(\omega),
\label{3.1}
\end{equation}
where
\begin{equation}
\lambda_j(x) = \left\{
\begin{array}{ll}
1 &  {\rm if\ }  x_{j-1} < x < x_j ,
\\
0 & {\rm otherwise}
\end{array}
\right.
\label{3.1a}
\end{equation}
is the characteristic function of the $j$th slab ($x_0\!\to\!-\infty$,
$x_N\!\to\!\infty$).
In particular, for $N\!\ge\!3$ the system represents an ($N\!-\!2$)-slab
dielectric plate surrounded by dielectric matter whose permittivity on the
left and right, respectively, is $\epsilon_1(\omega)$ and
$\epsilon_N(\omega)$. To determine the Green function $G(x,x',\omega)$ that
satisfies Eq.~(\ref{2.9}) [and vanishes at infinity], we note that
$G(x,x',\omega)$ can be decomposed into two parts,
\begin{equation}
G(x,x',\omega) = G_0(x,x',\omega) + G_1 (x,x',\omega),
\label{3.2}
\end{equation}
where, according to Eq.~(\ref{2.10a}),
\begin{equation}
G_0(x,x',\omega) = \sum_{j=1}^N \lambda_j(x) \lambda_j(x')
\left[2i\frac{\omega}{c} n_j (\omega)\right]^{-1}
\exp{\left[i \frac{\omega}{c} n_j (\omega) \left| x - x' \right|\right]},
\label{3.3}
\end{equation}
and $G_1(x,x',\omega)$ is solution of the homogeneous equation
\begin{equation}
\left[ \frac{\partial^2}{\partial x^2}
+ \frac{\omega^2}{c^2} \epsilon(x,\omega) \right] G_1 (x,x',\omega) = 0,
\label{3.4}
\end{equation}
which implies that
\begin{equation}
     G_1(x,x',\omega)
=
\sum_{j=1}^N \lambda_j(x) \left[
     C_j^{(1)}(x',\omega) e^{i n_j(\omega) \omega x/c}
+ C_j^{(2)}(x',\omega) e^{-i n_j(\omega) \omega x/c} \right]\!.
\label{3.4a}
\end{equation}
Clearly, the $C_j^{(1)}(x',\omega)$ and $C_j^{(2)}(x',\omega)$
must be determined in such a way that
$G(x,x',\omega)$ vanishes at infinity and is continuously differentiable
at the surfaces of discontinuity. The somewhat lengthy calculations may
be performed in a straightforward way. For the simplest case of a
single-slab dielectric plate embedded in dielectric matter ($N\!=\!3$,
cf. Fig.~1),
the result is given in
\ref{Gr}. Because of the
voluminous formulas, we renounce their presentation for the general case.

Combining Eqs.~(\ref{2.8}) and (\ref{3.2}) [together with Eqs.~(\ref{3.3})
and (\ref{3.4a})], the (Schr\"{o}dinger) operator of the vector potential
$\hat{A}(x)$ for the $j$th domain ($j\!=\!1,\ldots,N$) may be represented
as, similar to Eq.~(\ref{2.18}),
\begin{equation}
\hspace*{-2ex}
\hat{A} (x) = \int_0^{\infty} {\rm d} \omega \,
\sqrt{\frac{\hbar}{4 \pi c \omega \epsilon_0 \beta_j(\omega) {\cal A}}}
\,\frac{\beta_j(\omega)} {n_j(\omega)}
\left[
e^{i\beta_j(\omega) \omega x/c} \hat{a}_{j+}(x,\omega)
+ e^{-i\beta_j(\omega) \omega x/c} \hat{a}_{j-}(x,\omega)
\right] + {\rm H.c.}
\label{3.7}
\end{equation}
($x_{j-1}\!\leq\!x\!\leq\!x_j$),
where the dependence on $x$ of the amplitude operators
$\hat{a}_{j\pm}(x,\omega)$ is governed by quantum Langevin equations
of the type given in Eq.~(\ref{2.23}) together with Eqs.~(\ref{2.24})
and (\ref{2.25}) [$\beta(\omega),\gamma(\omega)\to
\beta_j(\omega),\gamma_j(\omega)$], so that $\hat{a}_{j\pm}(x,\omega)$
can be represented in the form (\ref{2.27}), viz.
\begin{eqnarray}
\lefteqn{
\hat{a}_{j\pm} (x,\omega) = \hat{a}_{j\pm} (x',\omega)
e^{\mp \gamma_j(\omega) \omega (x - x') / c}
}
\nonumber
\\ & &
\hspace{9.3ex}
+ \int_{x'}^x {\rm d} y\, \hat{F}_{j\pm} (y, \omega)
e^{\mp \gamma_j(\omega) \omega (x - y)/c}
\label{3.7a}
\end{eqnarray}
($x_{j-1}\!\leq\!x,x'\!\leq\!x_j$),
where
\begin{equation}
\hat{F}_{j\pm} (x, \omega) = \pm \frac{1}{i} \sqrt{2 \gamma_j(\omega)
\frac{\omega}{c}} \,
e^{\mp i \beta_j(\omega) \omega x/c}
\hat{f} (x, \omega).
\label{3.7b}
\end{equation}
Since the relations between the $\hat{a}_{j\pm}(x,\omega)$
and $\hat{f}(x,\omega)$ sensitively depend on the actual expression for
$G(x,x',\omega)$, the commutation relations (\ref{2.26})
[that are based on Eq.~(\ref{2.20})] cannot be applied in general.
Successive application of Eq.~({\ref{3.7a}}) enables one to relate
the (amplitude) operators $\hat{a}_{1-}(x,\omega)$
($-\infty\!\leq\!x\!\leq\!x_1$) and
$\hat{a}_{N+}(x,\omega)$ ($x_{N-1}\!\leq\!x\!\leq\!\infty$)
for the outgoing fields to the left and right, respectively,
to the operators of the corresponding incoming fields,
$\hat{a}_{1+}(x,\omega)$ and $\hat{a}_{N-}(x,\omega)$, and
operator noise sources associated with the losses owing to absorption.

\subsection{Single-slab dielectric plate}
\label{singleslab}
To illustrate the procedure outlined, let us first study the input--output
relations for a single-slab dielectric plate of thickness $l$
embedded in two (different) dielectrics ($N\!=\!3$, cf.
Fig.~1).
Substituting in Eq.~(\ref{2.8}) for $G(x,x',\omega)$ the
expression (\ref{a.4}) given in
\ref{Gr}
and introducing the (amplitude) operators
$\hat{a}_{1\pm}(x,\omega)$, $-\infty\!\leq\!x\!\leq\!-l/2$, and
$\hat{a}_{3\pm}(x,\omega)$, $l/2\!\leq\!x\!\leq\!\infty$
[\ref{expl}, Eqs.~(\ref{3.10a}) -- (\ref{3.10e})],
the input operators are found to satisfy the commutation relations
\begin{equation}
\big[ \hat{a}_{1+}(x, \omega), \hat{a}_{1+}^{\dagger} (x', \omega')
\big]  =  e^{-\gamma_1(\omega)|x-x'|/c} \delta(\omega-\omega'),
\label{3.10i}
\end{equation}
\begin{equation}
\big[ \hat{a}_{3-}(x, \omega), \hat{a}_{3-}^{\dagger} (x', \omega')
\big]  =  e^{-\gamma_3(\omega)|x-x'|/c} \delta(\omega-\omega'),
\label{3.10j}
\end{equation}
\begin{equation}
\big[ \hat{a}_{1+}(x, \omega), \hat{a}_{3-}^{\dagger} (x', \omega')
\big]  =  0
\label{3.10f}
\end{equation}
[Eqs.~(\ref{a.4d}) -- (\ref{a.4a})], so that the input
fields from the left and right behave like the fields
in the corresponding bulk dielectrics and may be regarded as
independent variables.
Note that $\hat{a}_{1+}(x,\omega),\,\hat{a}_{3-}(x,\omega)$ are
commuting quantities.

Defining operators
\begin{equation}
\hat{g}_\pm^{(1)}(\omega) = \left[ 2 c_\pm(l,\omega)
                                           \right]^{- \frac{1}{2} }
\left[
\hat{g}'_-(\omega)
\pm \hat{g}'_+(\omega) \right]\!,
\label{3.40}
\end{equation}
where
\begin{equation}
\hat{g}'_\pm(\omega)
= \frac{1}{i} \sqrt{\frac{\omega}{c}} \,
e^{i n_2 (\omega) \omega l/(2c)}
\int_{-\frac{l}{2}}^{\frac{l}{2}} {\rm d} x' \,
e^{\mp i n_2 (\omega) \omega x'/c} \hat{f} (x',\omega)
\label{3.10}
\end{equation}
and
\begin{equation}
c_\pm(l,\omega)
=
e^{-\gamma_2(\omega) \omega l/c}
\frac{1}{\gamma_2(\omega)}\,
\sinh{\!\left[
\gamma_2 (\omega) \frac{\omega}{c} l \right]}
\pm e^{-\gamma_2(\omega) \omega l/c}
\frac{1}{\beta_2 (\omega)}\,
\sin{\! \left[
\beta_2 (\omega) \frac{\omega}{c} l \right]}\!,
\label{3.17a}
\end{equation}
and recalling  Eqs.~(\ref{2.16}) and (\ref{2.17}), we find that
\begin{equation}
\left[
\hat{g}_{\pm}^{(1)} (\omega) ,
    \big(\hat{g}_{\pm}^{(1)} (\omega') \big)^{\dagger} \right]
= \delta(\omega - \omega') ,
\label{3.42}
\end{equation}
\begin{equation}
\left[
\hat{g}_{\pm}^{(1)} (\omega) ,
    \big(\hat{g}_{\mp}^{(1)} (\omega') \big)^{\dagger} \right] = 0 .
\label{3.43}
\end{equation}
Since
\begin{equation}
\left[ \hat{a}_{1+}(x, \omega),
   \big(\hat{g}_{\pm}^{(1)} (\omega') \big)^{\dagger}
\right] = 0 = \left[ \hat{a}_{3-}(x', \omega),
   \big(\hat{g}_{\pm}^{(1)} (\omega') \big)^{\dagger} \right]
\label{3.10g}
\end{equation}
[Eq.~(\ref{a.4b})], the incoming-field (amplitude) operators and
the operators $\hat{g}_\pm^{(1)}(\omega),\,\big(\hat{g}_\pm^{(1)}\big)
^\dagger(\omega)$ may be regarded as being
independent variables (note that $\hat{a}_{1+}$, $\hat{a}_{3-}$
and $\hat{g}_\pm^{(1)}$ commute). Moreover, the
$\hat{g}_\pm^{(1)}(\omega)$ and $\big(\hat{g}_\pm^{(1)}\big)
^\dagger(\omega)$, respectively, which are obviously destruction and
creation operators of bosonic excitations associated with the plate,
play the role of the additional operator noise sources in the
input--output relations
\begin{equation}
\left(\!\!\!
\begin{array}{c}
\hat{a}_{1-}\big(-\textstyle{\frac{1}{2}}l,\omega\big)
\\[.5ex]
\hat{a}_{3+}\big(\textstyle{\frac{1}{2}}l,\omega\big)
\end{array}
\!\!\!\right) \! = \! \widetilde{\bf T}^{(1)} \!\left(\!\!\!
\begin{array}{c}
\hat{a}_{1+}\big(-\textstyle{\frac{1}{2}}l,\omega\big)
\\[.5ex]
\hat{a}_{3-}\big(\textstyle{\frac{1}{2}}l,\omega\big)
\end{array}
\!\!\!\right) \! + \! \widetilde{\bf A}^{(1)}\! \left(\!\!
\begin{array}{c}
\hat{g}_{+}^{(1)}(\omega)
\\[.5ex]
\hat{g}_{-}^{(1)}(\omega)
\end{array}
\!\!\right)
\label{3.9}
\end{equation}
derived in
\ref{expl} [Eq.~(\ref{a.4c})], the elements of
the $2\times 2$ matrices
\begin{equation}
\widetilde{\bf T}^{(1)}  =
\left(
\begin{array}{cc}
 T_{11}^{(1)}(\omega) & T_{12}^{(1)}(\omega)
\\[.5ex]
T_{21}^{(1)}(\omega) & T_{22}^{(1)}(\omega)
\end{array}
\right)\!,
\label{3.11}
\end{equation}
and
\begin{equation}
\widetilde{\bf A}^{(1)} =
\left(
\begin{array}{cc}
 A_{11}^{(1)}(\omega) & A_{12}^{(1)}(\omega)
\\[.5ex]
A_{21}^{(1)}(\omega) & A_{22}^{(1)}(\omega)
\end{array}
\right)
\label{3.16}
\end{equation}
being given in Eqs.(\ref{a.5a}) -- (\ref{a.12a})
[note the simplifications (\ref{a.3a}) -- (\ref{b.4}) when
the plate is surrounded by vacuum].
In Eq.~(\ref{3.9}) the (amplitude) operators of the outgoing fields,
$\hat{a}_{1-}(-l/2,\omega)$ and $\hat{a}_{3+}(l/2,\omega)$,
are expressed in terms of the operators of the incoming fields,
$\hat{a}_{1+}(-l/2,\omega)$ and $\hat{a}_{3-}(l/2,\omega)$,
and the operator noise sources $\hat{g}_\pm^{(1)}(\omega)$.
The characteristic transformation matrix of the plate,
$ \widetilde{\bf T}^{(1)} $,
which for an approximately lossless dielectric plate in free space reduces
to the well-known characteristic matrix given, e.g., in \cite{BW},
describes the effects of transmission and reflection of the input fields,
whereas the losses inside the plate give rise to an additional matrix,
$ \widetilde{\bf A}^{(1)} $, which may be called characteristic
absorption matrix.

It should be mentioned that the output amplitude operators
$\hat{a}_{1-}(x)$, $x\!\leq\!-l/2$, and $\hat{a}_{3+}(x)$,
$x\!\geq\!l/2$, can easily be obtained from Eq.~(\ref{3.7a}),
with $x'\!=\!-l/2$ and $x'\!=\!l/2$, respectively, and application
of the input--output relations (\ref{3.9}). The resulting
representation of the outgoing fields is of course fully equivalent
to the Green's function expansion primarily used.
The commutation relations for the output amplitude operators are given in
\ref{expl} [Eqs.~(\ref{3.17}) -- (\ref{3.18})]. They differ,
in general, from those given in Eqs.~(\ref{3.10i}) -- (\ref{3.10f})
for the input operators. The differences vanish when the distances
from the plate are large compared with the (classical) absorption
length or when the plate is in free space.

\subsection{Multilayer dielectric plates}
\label{multislab}
The results given in Sec.~\ref{singleslab} can be extended to an arbitrary
multi-slab dielectric structure ($N\!\geq\!3$, cf.
Fig.~2)
in a straightforward way (for details see
\ref{induc}). In particular,
the input--output relations may be given by
\begin{equation}
\left(\!\!
\begin{array}{c}
\hat{a}_{1-}\big(x_1,\omega\big)
\\[.5ex]
\hat{a}_{N+}\big(x_{N-1},\omega\big)
\end{array}
\!\!\right) = \widetilde{\bf T}^{(N-2)} \left(\!\!
\begin{array}{c}
\hat{a}_{1+}\big(x_1,\omega\big)
\\[.5ex]
\hat{a}_{N-}\big(x_{N-1},\omega\big)
\end{array}
\!\!\right)
 +  \widetilde{\bf A}^{(N-2)} \left(\!\!
\begin{array}{c}
\hat{g}_{+}^{(N-2)}(\omega)
\\[.5ex]
\hat{g}_{-}^{(N-2)}(\omega)
\end{array}
\!\!\right)\!.
\label{3.27}
\end{equation}
Here, $x\!=\!x_1$ and $x\!=\!x_{N-1}$,
respectively, are the left and right surfaces of the multi-slab plate
(note that for the single-slab plate, $N\!=\!3$, the notations
$x_1\!=\!-l/2$ and $x_{N-1}\!=\!x_2\!=\!l/2$ have been used).
The commutation rules for the input operators
$\hat{a}_{1+}(x,\omega)$, $\hat{a}_{1+}^\dagger(x,\omega)$
($-\infty\!\leq\!x\!\leq\!x_1$),
$\hat{a}_{N-}(x,\omega)$, $\hat{a}_{N-}^\dagger(x,\omega)$
($x_{N-1}\!\leq\!x\!\leq\!\infty$), and the noise operators
$\hat{g}_{\pm}^{(N-2)}(\omega)$ are the same as in the preceding
section, i.e.,
\begin{equation}
\big[ \hat{a}_{1+}(x, \omega), \hat{a}_{1+}^{\dagger} (x', \omega')
\big]  =  e^{-\gamma_1(\omega)|x-x'|/c} \delta(\omega-\omega'),
\label{3.27i}
\end{equation}
\begin{equation}
\big[ \hat{a}_{N-}(x, \omega), \hat{a}_{N-}^{\dagger} (x', \omega')
\big]  =  e^{-\gamma_N(\omega)|x-x'|/c} \delta(\omega-\omega'),
\label{3.27j}
\end{equation}
\begin{equation}
\big[ \hat{a}_{1+}(x, \omega), \hat{a}_{N-}^{\dagger} (x', \omega')
\big]  =  0,
\label{3.27f}
\end{equation}
\begin{equation}
\left[
\hat{g}_{\pm}^{(N-2)} (\omega) ,
            \big( \hat{g}_{\pm}^{(N-2)} (\omega') \big)^{\dagger}
\right] = \delta(\omega - \omega') ,
\label{3.27g}
\end{equation}
\begin{equation}
\left[
\hat{g}_{\pm}^{(N-2)} (\omega) ,
            \big( \hat{g}_{\mp}^{(N-2)} (\omega') \big)^{\dagger}
\right]  =  0,
\label{3.27h}
\end{equation}
\begin{equation}
\left[ \hat{a}_{1+}(x, \omega),
   \big(\hat{g}_{\pm}^{(N-2)} (\omega') \big)^{\dagger}
\right] = 0
 = \left[ \hat{a}_{N-}(x', \omega),
   \big(\hat{g}_{\pm}^{(N-2)} (\omega') \big)^{\dagger} \right]\!.
\label{3.27k}
\end{equation}

The input-output relations (\ref{3.27}) [together with
the commutation relations (\ref{3.27i}) -- (\ref{3.27k})] apply to
arbitrary multi-slab dielectric equipments described in terms
of a complex permittivity that spatially varies as a multi-step function
and whose dependence on frequency is consistent with the
Kramers--Kronig relations over the whole frequency domain.
Typical examples are fractionally transparent dielectric mirrors and
combinations of them, such as resonator-like cavities bounded by
dielectric walls. In particular, when the overall device is
surrounded by vacuum, so that the incoming and outgoing radiation fields
propagate in free space,
\begin{equation}
     n_1(\omega) = n_N(\omega) \equiv 1,
\label{3.16a}
\end{equation}
the familiar mode expansions for the input and output (free) fields
are recognized. For $j\!=\!1,N$ Eq.~(\ref{3.7}) takes the form
\begin{equation}
\hat{A} (x) = \int_0^{\infty} {\rm d} \omega \,
\sqrt{\frac{\hbar}{4 \pi c \omega \epsilon_0 {\cal A}}}
\left[
e^{i\omega x/c} \hat{a}_{j+}(\omega)
+ e^{-i\omega x/c} \hat{a}_{j-}(\omega)
\right] + {\rm H.c.},
\label{3.16b}
\end{equation}
where the input and output operators
$\hat{a}_{1\pm}(\omega)$  [$\hat{a}_{1\pm}^\dagger(\omega)$] and
$\hat{a}_{N\pm}(\omega)$ [$\hat{a}_{N\pm}^\dagger(\omega)$], respectively,
are proper (space-independent) photon destruction [creation] operators, and
\begin{equation}
\left(
\begin{array}{c}
\hat{a}_{1-}(\omega)
\\[.5ex]
\hat{a}_{N+}(\omega)
\end{array}
\right)  = \widetilde{\bf T}^{(N-2)} \left(
\begin{array}{c}
\hat{a}_{1+}(\omega)
\\[.5ex]
\hat{a}_{N-}(\omega)
\end{array}
\right)
+ \widetilde{\bf A}^{(N-2)} \left(\!
\begin{array}{c}
\hat{g}_{+}^{(N-2)}(\omega)
\\[.5ex]
\hat{g}_{-}^{(N-2)}(\omega)
\end{array}
\!\right)\!.
\label{3.16c}
\end{equation}
The influence of the plate on the incident light through
reflection and transmission from the two sides are described by the
matrix elements $T_{ik}^{(N-2)}(\omega)$, whereas the
matrix elements $A_{ik}^{(N-2)}(\omega)$ arise from absorption.
 From Eqs.~(\ref{3.27i}) -- (\ref{3.27f}), the
bosonic commutation relations for the input-mode operators
$\hat{a}_{1+}(\omega)$, $\hat{a}_{N-}(\omega)$
are easily seen to be satisfied.
Using them and recalling the commutation rules (\ref{3.27g}) --
(\ref{3.27k}), the bosonic commutation relations for the output-mode
operators $\hat{a}_{1-}(\omega)$, $\hat{a}_{N+}(\omega)$
can then be obtained by means of the input--output relations
(\ref{3.16c}), because the relations
\begin{eqnarray}
\lefteqn{
\big|T_{11}^{(N-2)}\big|^2 + \big|T_{12}^{(N-2)}\big|^2
+ \big|A_{11}^{(N-2)}\big|^2 + \big|A_{12}^{(N-2)}\big|^2
}
\nonumber \\
&&
= \big|T_{21}^{(N-2)}\big|^2 + \big|T_{22}^{(N-2)}\big|^2
+ \big|A_{21}^{(N-2)}\big|^2 + \big|A_{22}^{(N-2)}\big|^2
= 1
\nonumber \\
&&
\label{3.25}
\end{eqnarray}
and
\begin{eqnarray}
\lefteqn{
T_{11}^{(N-2)} \big(T_{21}^{(N-2)}\big)^*
+ T_{12}^{(N-2)} \big(T_{22}^{(N-2)}\big)^*
}
\nonumber \\
&& \hspace{2ex}
+ A_{11}^{(N-2)} \big(A_{21}^{(N-2)}\big)^*
+ A_{12}^{(N-2)} \big(A_{22}^{(N-2)}\big)^*  = 0
\label{3.26a}
\end{eqnarray}
are valid [see
\ref{expl} and \ref{induc}]. For notational
convenience, the frequency argument of the matrix elements
$T_{ik}^{(N-2)}$ and $A_{ik}^{(N-2)}$ matrices are omitted.
In particular, these relations reflect the fact that the sum of
the probabilities for reflection, transmission, and absorption of a
photon is equal to unity. If the losses inside the plate can
approximately be disregarded, $\widetilde{\bf A}^{(N-2)}\!\approx\!0$,
the well-known method of unitary transformation is recognized.
In this case the relations (\ref{3.25}) and (\ref{3.26a}) simplify to
\begin{equation}
\big|T_{11}^{(N-2)}\big|^2 + \big|T_{12}^{(N-2)}\big|^2
           = \big|T_{21}^{(N-2)}\big|^2 + \big|T_{22}^{(N-2)}\big|^2  = 1,
\label{3.16d}
\end{equation}
\begin{equation}
T_{11}^{(N-2)} \big(T_{21}^{(N-2)}\big)^\ast
             + T_{12}^{(N-2)} \big(T_{22}^{(N-2)}\big)^\ast = 0,
\label{3.16e}
\end{equation}
so that $\widetilde{\bf T}^{(N-2)}$ becomes a unitary matrix.
Since the photon operators of the output and input modes are
uniquely related to each other through a unitary transformation,
the bosonic commutation relations are automatically preserved.
In general, the $\widetilde{\bf T}^{(N-2)}$
matrix is not unitary and the output-mode operators are obtained,
according to Eq.~(\ref{3.16c}), from both the input-mode operators and the
noise operators associated with the losses. The relations
(\ref{3.25}) and (\ref{3.26a}), which are the natural generalization of
the relations (\ref{3.16d}) and (\ref{3.16e}), may be regarded
as necessary conditions imposed on the $\widetilde{\bf T}^{(N-2)}$ and
$\widetilde{\bf A}^{(N-2)}$ matrices of an arbitrary dispersive and
absorptive multi-slab dielectric device in free space. It should
be emphasized that these conditions need not be postulated, but
they necessarily come out of the theory, which also enables one
to systematically calculate both the $\widetilde{\bf T}^{(N-2)}$
and $\widetilde{\bf A}^{(N-2)}$ matrices.

\section{Radiation-field correlation functions}
\label{corrfunc}
The input--output relations (\ref{3.16c}) can be used
to obtain the quantum statistical properties of the outgoing
radiation from the properties of the incoming radiation and the
excitations associated with the dielectric matter.
With regard to measurement, the quantum statistics of radiation
is frequently described in terms of normally ordered correlation
functions, such as correlation functions of the electric-field
strength or the photon creation and destruction operators
(see, e.g., \cite{3a,Louise}).
Introducing the notations
$\hat{a}_1\!\equiv\!\hat{a}_{1+}$, $\hat{a}_2\!\equiv\!\hat{a}_{N-}$,
$\hat{a}'_1\!\equiv\!\hat{a}_{1-}$, $\hat{a}'_2\!\equiv\!\hat{a}_{N+}$
and $\hat{g}_1\!\equiv\!\hat{g}_{+}^{(N-2)}$,
$\hat{g}_2\!\equiv\!\hat{g}_{-}^{(N-2)}$,
from Eqs.~(\ref{2.12}) and (\ref{3.16b}), the electric-field strength
of the outgoing radiation in the $i$th channel ($i\!=\!1,2$)
reads as
\begin{equation}
    \hat{E}'_i(x) =  \hat{E}'\,\!_{\!\! i}^{(+)}(x)
                  +  \hat{E}'\,\!_{\!\! i}^{(-)}(x),
\label{3.28a}
\end{equation}
\begin{equation}
\hat{E}'\,\!_{\!\! i}^{(+)}(x) = i \int_0^{\infty} {\rm d} \omega \,
\sqrt{\frac{\hbar \omega}{4 \pi c \epsilon_0 {\cal A}}}
e^{i\omega \eta_i x/c} \hat{a}'_{i}(\omega),
\label{3.28a1}
\end{equation}
\begin{equation}
\hat{E}'\,\!_{\!\! i}^{(-)}(x) =
            \left[ \hat{E}'\,\!_{\!\! i}^{(+)}(x) \right]^{\dagger}
\label{3.28a2}
\end{equation}
($\eta_1\!=-1$, $\eta_2\!=\!1$), where the output-photon operators
$\hat{a}_i'(\omega)$ can be related to the input-photon operators
$\hat{a}_i(\omega)$ as, according to Eq.~(\ref{3.16c}),
\begin{equation}
\hat{a}'_i(\omega)  = \sum_{k=1}^2 \big[
           T_{ik}(\omega)\, \hat{a}_{k}(\omega)
                     + A_{ik}(\omega)\, \hat{g}_{k}(\omega) \big]
\label{3.28b}
\end{equation}
[$T_{ik}(\omega)\!\equiv\!T_{ik}^{(N-2)}(\omega)$,
$A_{ik}(\omega)\!\equiv\!A_{ik}^{(N-2)}(\omega)$).
To express the normally ordered electric-field correlation
functions of the outgoing radiation,
\begin{equation}
{C'}_{\!\!\{i_\mu\}}^{(m,n)} \! \left( \big\{ x_{\mu},t_{\mu} \big\} \right)
 = \left\langle
\left[
\prod\limits_{\mu=1}^m
     \hat{E}'\,\!_{\!\! i_\mu}^{(-)} \left( x_\mu,t_\mu \right)
\right]
\left[
\prod\limits_{\mu=m+1}^{m+n}
     \hat{E}'\,\!_{\!\! i_\mu}^{(+)} \left( x_\mu,t_\mu \right)
\right]
\right\rangle\!,
\label{3.28}
\end{equation}
in terms of normally ordered correlation functions of photon creation
and destruction operators, we use Eqs.~(\ref{3.28a}) -- (\ref{3.28a2})
and recall the harmonic (exponential) time evolution of the photon
destruction operators in the Heisenberg picture. We obtain
\begin{eqnarray}
\lefteqn{
{C'}_{\!\!\{i_\mu\}}^{(m,n)} \! \left( \big\{ x_{\mu},t_{\mu} \big\} \right)
=  i^{n-m}
\left(\frac{\hbar}{4 \pi c \epsilon_0 {\cal A}}\right)^{\!\frac{n+m}{2}}
}
\nonumber \\
&& \hspace{2ex}
\int_0^{\infty} {\rm d} \omega_1 \, \sqrt{\omega_1}
   e^{i\omega_1\tau_{i_1}}
\ldots
\int_0^{\infty} {\rm d} \omega_{m+n} \, \sqrt{\omega_{m+n}}
  e^{-i\omega_{m+n}\tau_{i_{m+n}}}
{\underline{C}'}_{\{i_\mu\}}^{(m,n)} \left( \big\{ \omega_\mu \big\}
\right)\!
\label{3.32}
\end{eqnarray}
($\tau_{i_\mu}\!=\!t_\mu\!+\!\eta_{i_\mu} x_\mu/c$), where
\begin{equation}
{\underline{C}'}_{\{i_\mu\}}^{(m,n)} \left( \big\{ \omega_\mu \big\} \right)
= \left\langle
\left[
\prod\limits_{\mu=1}^m
\hat{a}'\,\!_{\!\! i_\mu}^{\dagger}\! \left( \omega_\mu \right)
\right]
\left[
\prod\limits_{\mu=m+1}^{m+n}
\hat{a}'_{i_\mu}\! \left( \omega_\mu \right)
\right]
\right\rangle\!,
\label{3.32a}
\end{equation}
which can be rewritten as, on using the input--output relations
(\ref{3.28b}), Eq.~(\ref{3.32a}),
\begin{eqnarray}
\lefteqn{
{\underline{C}'}_{\{i_\mu\}}^{(m,n)} \!
\left( \big\{ \omega_\mu \big\} \right)
= \left\langle
\left\{
\prod\limits_{\mu=1}^m
\sum_{k_\mu=1}^2 \left[ T_{i_\mu k_\mu}^\ast(\omega_\mu)\,
                     \hat{a}_{k_\mu}^{\dagger}(\omega_\mu)
                  + A_{i_\mu k_\mu}^\ast(\omega_\mu)\,
                     \hat{g}_{k_\mu}^\dagger(\omega_\mu) \right]
\right\}
\right.  }
\nonumber \\
&&
\hspace{10ex} \times  \left.
\left\{
\prod\limits_{\mu=m+1}^{m+n}
\sum_{k_\mu=1}^2 \left[ T_{i_\mu k_\mu}(\omega_\mu)\,
                     \hat{a}_{k_\mu}(\omega_\mu)
                  + A_{i_\mu k_\mu}(\omega_\mu)\,
                     \hat{g}_{k_\mu}(\omega_\mu) \right]
\right\}
\right\rangle\!.
\label{3.32b}
\end{eqnarray}
In particular, when the states of the incoming radiation and the
dielectric matter are not correlated to each other, the correlation
functions of the output photons, Eq.~(\ref{3.32b}), can be expressed
in terms of sums of products of input-photon correlation functions
\begin{equation}
{\underline{C}}_{\{i_\mu\}}^{(k,l)} \!
\left( \big\{ \omega_\mu \big\} \right)
= \left\langle
\left[
\prod\limits_{\mu=1}^k
\hat{a}_{i_\mu}^{\dagger}\! \left( \omega_\mu \right)
\right]
\left[
\prod\limits_{\mu=l+1}^{k+l}
\hat{a}_{i_\mu}\! \left( \omega_\mu \right)
\right]
\right\rangle
\label{3.32c}
\end{equation}
and correlation functions of the excitations associated with the
dielectric matter,
\begin{equation}
{\underline{\Gamma}}_{\{i_\mu\}}^{(p,q)} \!
\left( \big\{ \omega_\mu \big\} \right)
= \left\langle
\left[
\prod\limits_{\mu=1}^p
\hat{g}_{i_\mu}^{\dagger}\! \left( \omega_\mu \right)
\right]
\left[
\prod\limits_{\mu=p+1}^{p+q}
\hat{g}_{i_\mu}\! \left( \omega_\mu \right)
\right]
\right\rangle\!,
\label{3.32d}
\end{equation}
with $k,p\!\leq\!m$ and $l,q\!\leq\!n$. Clearly, when the matter
is prepared in an incoherent state, then the correlation functions
${\underline{\Gamma}}_{\{i_\mu\}}^{(p,q)}(\{ \omega_\mu \})$ vanish
when $p\!\neq\!q$
(explicit expressions for the correlation functions observed when
the matter is thermally excited are given in
\ref{illum}).

To give a simple example, let us consider the photon-number
density (number of photons per unit frequency) in the $i$th output
channel ($i\!=\!1,2$),
\begin{equation}
{N}'_{{\rm ph}\,i} (\omega)
= \left\langle \hat{a}'\,\!_{\!\! i}^{\dagger} (\omega)
               \hat{a}'_{i} (\omega)\right\rangle
= {\underline{C}'}_{ii}^{(1,1)} \!
              \left( \omega,\omega \right).
\label{3.32e}
\end{equation}
Using Eq.~(\ref{3.32b}) and assuming that the overall state
is factored, we find that
\begin{eqnarray}
N'_{{\rm ph}\,i} (\omega) & = & \sum_{k=1}^2 \left[
\left|T_{ik}(\omega)\right|^2 N_{{\rm ph}\,k} (\omega)
+ \left|A_{ik}(\omega)\right|^2 N_{{\rm dp}\,k} (\omega) \right]
\nonumber \\
& &
\hspace{-9ex}
+ \, \left[
  T_{i1}^\ast(\omega)
     T_{i2}(\omega)
      \left\langle \hat{a}_1^\dagger (\omega)
                                 \hat{a}_2 (\omega) \right\rangle
+ A_{i1}^\ast(\omega)
     A_{i2}(\omega)
           \left\langle \hat{g}_1^\dagger (\omega)
                                 \hat{g}_2 (\omega) \right\rangle
+ {\rm c.c.} \right]\!,
\label{3.32f}
\end{eqnarray}
where
\begin{equation}
{N}_{{\rm ph}\,k} (\omega)
= \left\langle \hat{a}_{k}^{\dagger} (\omega)
               \hat{a}_{k} (\omega)\right\rangle
\label{3.32g}
\end{equation}
and
\begin{equation}
{N}_{{\rm dp}\,k} (\omega)
= \left\langle \hat{g}_{k}^{\dagger} (\omega)
               \hat{g}_{k} (\omega)\right\rangle
\label{3.32h}
\end{equation}
($k\!=\!1,2$) are the number densities of the input photons and the
excitations associated with the dielectric matter, respectively.

Let us first consider the case when the input field is in the vacuum state,
\begin{equation}
\left\langle \hat{a}_{1}^\dagger (\omega)
      \hat{a}_{2} (\omega) \right\rangle = 0 ,
\label{3.32h1}
\end{equation}
\begin{equation}
      {N}_{{\rm ph}\,1} (\omega) = {N}_{{\rm ph}\,2} (\omega) = 0,
\label{3.32h1a}
\end{equation}
and the dielectric plate is in thermal equilibrium,
\begin{equation}
\left\langle \hat{g}_{1}^\dagger (\omega)
      \hat{g}_{2} (\omega) \right\rangle = 0 ,
\label{3.32h2}
\end{equation}
\begin{equation}
{N}_{{\rm dp}\,1} (\omega) \Delta\omega
= {N}_{{\rm dp}\,2} (\omega) \Delta\omega
      =\frac{{\cal L}}{2\pi c} {n}_{\rm th}(\omega) \Delta\omega,
\label{3.32h3}
\end{equation}
where
\begin{equation}
{n}_{\rm th}(\omega) =
\frac{1}{\exp\!\left(\frac{\hbar\omega}{k_{\rm B}T}\right) - 1}
\label{3.32h3a}
\end{equation}
($T$, temperature; $k_{\rm B}$, Boltzmann constant). For the sake of
convenience, in Eq.~(\ref{3.32h3}) we have assumed a large but finite
length ${\cal L}$ of the quantization volume of the radiation, so that
the frequency spectrum may be regarded as being discrete. Using
Eqs.~(\ref{3.32h1}) -- (\ref{3.32h3}), we see that
Eq.~(\ref{3.32f}) reduces to
\begin{equation}
N'_{{\rm ph}\,i} (\omega)
= \frac{{\cal L}}{2\pi c} \alpha_i(\omega) \, n_{\rm th} (\omega)
\qquad (i = 1,2),
\label{3.32h4}
\end{equation}
where
\begin{equation}
\alpha_i(\omega) =
\sum_{k=1}^2 \left|A_{ik}(\omega)\right|^2
\label{3.32h5}
\end{equation}
is the ($i$th-side) absorption coefficient of the plate.
The result (\ref{3.32h4}) is in full agreement with the standard
quantum theory of thermal radiation. The plate behaves
like a thermal radiator which tends to a black body as the
absorption becomes perfect [$\alpha_i(\omega)\!\to\!1$].
Note that for ${\cal L}\!\to\infty$ the photon numbers per unit
frequency and unit length, $N'_{\rm ph\,i}(\omega)/{\cal L}$, remain
finite. When the plate is in thermal equilibrium and the incident light is
black-body radiation,
\begin{equation}
{N}_{{\rm ph}\,1} (\omega) \Delta\omega
= {N}_{{\rm ph}\,2} (\omega) \Delta\omega
= \frac{{\cal L}}{2\pi c} {n}_{\rm th}(\omega) \Delta\omega,
\label{3.32h6}
\end{equation}
the outgoing light is also expected to be black-body radiation.
Indeed, combining Eqs.~(\ref{3.32h}), (\ref{3.32h1}), (\ref{3.32h2}),
using Eqs.~(\ref{3.32h3}), (\ref{3.32h6}), and recalling the relations
(\ref{3.25}), we find that
\begin{equation}
N'_{{\rm ph}\,i} (\omega)
= \frac{{\cal L}}{2\pi c} n_{\rm th} (\omega)
\qquad (i = 1,2).
\label{3.32h7}
\end{equation}

Finally, let us consider the case when a zero-tem\-per\-ature dielectric
plate is irradiated from one side,
\begin{equation}
      N_{{\rm ph}\,1} (\omega) \geq 0, \quad
      N_{{\rm ph}\,2} (\omega) = 0,
\label{3.32h8}
\end{equation}
\begin{equation}
      N_{{\rm dp}\,1} (\omega) = N_{{\rm dp}\,2} (\omega) = 0.
\label{3.32h9}
\end{equation}
 From Eq.~(\ref{3.32f}) we obtain, on using Eqs.~(\ref{3.32h1}) and
(\ref{3.32h2}) together with Eq.~(\ref{3.32h8}) and (\ref{3.32h9}),
\begin{equation}
N'_{{\rm ph}\,i} (\omega)
= \left| T_{i1} (\omega) \right|^2 N_{{\rm ph}\,1} (\omega)
\qquad (i = 1,2).
\label{3.32h10}
\end{equation}
The mean photon-number densities in the output channels 1 and 2, respectively,
are simply given by the mean input photon-number density multiplied by the
reflection and transmission coefficients. In general, the
overall output photon-number density is reduced below the input level
owing to absorption:
\begin{equation}
N'_{{\rm ph}\,1} (\omega) + N'_{{\rm ph}\,2} (\omega)
\leq N_{{\rm ph}\,1} (\omega).
\label{3.32h11}
\end{equation}

In Figs.~3 and 4 the relative photon-number densities
of the outgoing radiation, ${\cal N}_1$ $\equiv$ $N'_{{\rm ph}\,1}
(\omega)$\,$/$\,$N_{{\rm ph}\,1}(\omega)$ $=$ $|T_{11}(\omega)|^2$ and
${\cal N}_2$ $\equiv$ $N'_{{\rm ph}\,2}(\omega)$\,$/$\,$N_{{\rm ph}\,1}
(\omega)$ $=$ $|T_{21}(\omega)|^2$ [Eq.~(\ref{3.32h10})], are shown as
functions of frequency and plate thickness for a single-slab plate in
free space ($\epsilon_1(\omega)\!=\!\epsilon_3(\omega)\!=\!1$).
The relative photon-number density of the radiation absorbed by the
plate, ${\cal N}_{\rm a}$ $=$ $[N_{{\rm ph}\,1}
(\omega)$ $-$ $(N'_{{\rm ph}\,1}(\omega)$ $+$ $N'_{{\rm ph}\,2}
(\omega))]$\,$/$\,$N_{{\rm ph}\,1}(\omega)$ $=$ $\alpha_1
(\omega)$ $=$ $\alpha_2(\omega)$
is shown in
Fig.~5.
The results in
Figs.~3 -- 5
are given for a simple model permittivity
$\epsilon(\omega)\!\equiv\!\epsilon_2(\omega)$
based on a single medium resonance,
\begin{equation}
\epsilon(\omega) = 1+\frac{\omega_1^2}
{\omega_0^2 - \omega^2 - i \Gamma \omega},
\label{3.32q}
\end{equation}
which enables one to clearly distinguish the (resonance) region of
frequency for which the imaginary part of the refractive
index may substantially exceed the real part from other
(off-resonance) regions for which the imaginary part becomes small.

For frequencies that are small compared with the medium
resonance frequency ($\omega\!\ll\!\omega_0$) the approximately real
refractive index ($n(\omega)\!=\!\beta(\omega)\!+
\!i\gamma(\omega)\!\approx\!\beta(\omega)\!\geq\! 1$) gives rise to a
variation with the plate thickness of the
transmitted and reflected numbers of photons which show the
oscillating behaviour typical for a Fabry--P\'{e}rot device. In
this frequency region the losses inside the plate may be disregarded.
Increasing the frequency, both the real and imaginary parts of the
refractive index, $\beta(\omega)$ and $\gamma(\omega)$, respectively,
are increased. Increasing $\gamma(\omega)$ implies increasing
probability for photon absorption in the plate. Since the number
of absorbed photons increases with the thickness of the plate,
the number of transmitted photons decreases with increasing plate
thickness. Further, owing to the increasing absolute value of refractive
index the number of photons that are reflected are increased
at the expense of the number of photons that enter the plate.
The two effects mentioned become more and more pronounced as
$\omega$ approaches $\omega_0$. In particular, in the vicinity of
the medium resonance the number of reflected photons is substantially
enhanced. The photons that enter the plate are absorbed over a
short distance, so that the number of transmitted photons rapidly
tends to zero when the thickness of the plate is increased.
In this ``surface regime'' the plate behaves like a lossy mirror,
the enhanced reflectivity being caused by the large absolute
value of the refractive index, which results, in general, from both
the real and the imaginary parts. Further increase of frequency
that is associated with a decrease of the real and imaginary parts
of the refractive index (region of anomalous dispersion) reduces
the effects of strong reflection and absorption and the plate again
becomes fractionally transparent. Needless to say, that for
sufficiently high frequencies when the refractive index approaches
to unity the plate becomes fully transparent. The differences in the
response of the plate to the incoming radiation for the different
regions of frequency are less pronounced when the value of the damping
constant $\Gamma$ in Eq.~(\ref{3.32q}) is increased.

In practice, the permittivity is of course a much more complicated
function of frequency than the model function (\ref{3.32q}), because
of the multi-frequency resonance behaviour corresponding to the
multi-level excitation spectrum of the medium. In
Fig.~6 the real
and imaginary parts of the refractive index of amorphic SiO$_2$ are
plotted for a region of frequency including absorbing and nonabsorbing
sectors \cite{18}. To compare the results for this realistic
permittivity with those for the single resonance model (\ref{3.32q}), in
Fig.~7 the number of transmitted photons is shown as a function
of frequency and thickness of the plate. Although Fig.~7 roughly
agrees with Fig.~4, there are a number of differences in the details
which obviously result from the multi-frequency resonance structure.
We would like to mention that, on following the line in
\ref{induc}, the
characteristic transformation and absorption matrices can be calculated
numerically for an arbitrary $N$-slab device, as we will show in a forthcoming
paper.

\section{Summary and conclusions}
\label{sum}
Applying the method of Green-function expansion to the quantization
of radiation propagating through a multilayer dispersive and absorptive
dielectric plate, we have studied the problem of calculating the
proper input--output relations for the radiation field operators and
presented results for the case when the radiation propagates
perpendicularly to the plate. The plate is described in terms of a
multistep (spatially varying) complex permittivity in the frequency domain,
which is introduced phenomenologically and only required to satisfy
the Kramers--Kronig relations. The advantage of the method is that
it enables one to obtain input--output relations that do not only apply
to regions of frequency far from the medium resonances, but they are valid,
within the frame of the phenomenological linear electrodynamics, in the
whole frequency domain.

In consequence of the inclusion of the losses in the theory the
output-radiation field operators are found to be related to the
input-radiation field operators and operator noise sources in the
plate associated with the losses, in agreement with the
dissipation--fluctuation theorem. Disregarding all the losses,
the characteristic absorption matrix that relates the output-radiation
operators to the operators noise sources vanishes and the
characteristic transformation matrix that relates the output-radiation
operators to the input operators reduces to a unitary matrix.
The unitary transformation ensures that the bosonic commutation
relations are preserved.

When the multilayer dielectric plate is embedded in an absorbing medium
the input and output radiation fields can be described in terms of amplitude
operators whose space dependence (owing to damping) is governed
by quantum Langevin equations. Only in the limiting case when the
surrounding medium can be regarded as being lossless (particularly,
when the plate is embedded in free space), the amplitude operators
reduce to the well-known bosonic photon operators. In this case,
the characteristic transformation and absorption matrices of the plate
can be shown to be related to each other through conditions that
ensure preservation of the bosonic commutation relations. These
conditions can be regarded as the natural generalization of
the familiar unitarity conditions for lossless plates.

The input--output relations can advantageously be used in order
to obtain the quantum statistics of the output radiation from that
of the input radiation and the noise sources associated with the
absorbing matter. With regard to measurement, we have considered
normally ordered radiation field correlation functions.
To illustrate the theory, we have studied the input--output
relations for the (spectral) photon-number densities in more
detail. Restricting attention to a single-slab dielectric plate,
numerical results are presented for a single-resonance complex
(model) permittivity, and the applicability of the method to realistic
dielectric matter, such as SiO$_2$, is demonstrated.

Finally, let us briefly comment on the underlying formalism
of quantization of the phenomenological Maxwell theory for
radiation in dispersive and absorptive linear media. It should
be mentioned that the formalism used resembles the concepts of
generalized free-field theories \cite{Green}.
This type of quantum field theory has recently been applied successfully
to thermo field dynamics for quantum fields \cite{Hen}.
The similarities between the two descriptions may provide further insight
in the basic physical structure of the theory, such as the limit of
vanishing absorption. In thermo field dynamics this limit corresponds
to the zero-temperature limit that has been shown to be essentially
non-analytical.
Similar features are also found in an indeterminacy of the dispersion
relations. Absorption prevents the ``spatially damped photons'' from
exhibiting a well-determined relation between energy (frequency) and
momentum (wave vector).
Similarly, at finite temperature particle states achieve a continuous
spectrum not only for the particle momentum but at the same time
for the mass parameter (for any fixed value of the particle momentum).

\section*{Acknowledgement}
We would like to thank L. Kn\"oll for useful discussions.
One of us (T.G.) is grateful to P.D. Drummond for helpful comments on
the basic concepts of quantization and to P.A. Henning for the interesting
discussion on thermo fields.
This work was supported by the Deutsche Forschungsgemeinschaft
(74021\,40\,223).

\begin{appendix}
\renewcommand{\thesection}{Appendix \Alph{section}}
\section{Single-slab Green function}
\label{Gr}
\setcounter{equation}{0}
\renewcommand{\theequation}{\Alph{section} \arabic{equation}}
Let us consider a dielectric slab of thickness $l$ embedded in dielectric
matter that may be different on the two sides (see Fig.~1).
The Green function $G(x,x',\omega)$ can be obtained from Eqs.~(\ref{3.2}),
(\ref{3.3}), and (\ref{3.4a}), with $N\!=\!3$. The coefficients
$C_j^{(1)}(x',\omega)$ and $C_j^{(2)}(x',\omega)$
in Eq.~(\ref{3.4a}), $j\!=\!1,2,3$, must be determined from the
conditions that the Green function is continuously differentiable
at the surfaces of discontinuity (that is to say, at $x\!=\!\pm l/2$)
and vanishes at infinity. Note that the latter requires the coefficients
$C_1^{(1)}(x',\omega)$ and $C_3^{(2)}(x',\omega)$ to be zero.
Straightforward but rather lengthy calculation yields
[$n_j\!\equiv\!n_j(\omega)$]
\begin{eqnarray}
\lefteqn{
G(x,x')  = \left[ 2 i n_1 \frac{\omega}{c} \right]^{-1}
\Theta\!\left(- x' - {\textstyle{\frac{1}{2}}} l \right)
\left\{
\Theta\!\left( - x - {\textstyle{\frac{1}{2}}} l \right)
\Big[
e^{i n_1 \omega |x-x'| / c }
\right.
}
\nonumber
\\ & &
\left.
+ e^{i n_1 \omega | (-l)/2 - x'| / c }
\left(
r_{12} + t_{12} \vartheta e^{i n_2 \omega l/c} r_{23}
e^{i n_2 \omega l/c} t_{21}
\right)
e^{i n_1 \omega | (-l)/2 - x | /c }
\Big]
\right.
\nonumber
\\ & &
\left.
+ \left[
\Theta\!\left( x + {\textstyle{\frac{1}{2}}} l \right)
- \Theta\!\left( x - {\textstyle{\frac{1}{2}}} l \right)
\right]
t_{12} \vartheta e^{i n_1 \omega | (-l)/2 - x' | / c}
\left(
e^{i n_2 \omega | x - (-l)/2 | /c }
+ r_{23} e^{i n_2 \omega ( l + | l/2 - x | /c )}
\right)
\right.
\nonumber
\\ & &
\left.
+ \Theta\!\left( x - {\textstyle{\frac{1}{2}}} l \right)
e^{i n_1 \omega |  (-l)/2 - x' | / c}
t_{12} e^{i n_2 \omega l /c } \vartheta t_{23}
e^{i n_3 \omega | x - l/2 | / c}
\right\}
\nonumber
\\ & &
+ \left[ 2 i n_2 \frac{\omega}{c} \right]^{-1}
\left[
\Theta \left( x' + {\textstyle{\frac{1}{2}}} l \right)
- \Theta \left( x' - {\textstyle{\frac{1}{2}}} l \right)
\right]
\nonumber
\\ & &
\times
\left\{
\Theta \left( - x - {\textstyle{\frac{1}{2}}} l \right)
\vartheta
\left(
e^{i n_2 \omega | x' - (-l)/2 | / c}
+ e^{i n_2 \omega | l/2 - x' | / c} r_{23}
e^{i n_2 \omega l/c}
\right)
t_{21} e^{i n_1 \omega | (-l)/2 - x | / c}
\right.
\nonumber
\\ & &
\left.
+ \left[
\Theta\!\left( x  + {\textstyle{\frac{1}{2}}} l \right)
- \Theta\!\left( x - {\textstyle{\frac{1}{2}}} l \right)
\right]
\left[
e^{i n_2 \omega | x - x' |/c}
\right.
\right.
\nonumber
\\ & &
\left.
\left.
+  \vartheta
\left(
e^{i n_2 \omega | x' - (-l)/2 | / c}
r_{21} e^{i n_2 \omega l/c}
+ e^{i n_2 \omega | l/2 - x' | /c}
\right)
r_{23} e^{i n_2 \omega | l/2 - x | /c}
\right.
\right.
\nonumber
\\ & &
\left.
\left.
+ \vartheta
\left(
e^{i n_2 \omega | x' - (-l)/2 | /c}
+ e^{i n_2 \omega | l/2 - x' | /c } r_{23}
e^{i n_2 \omega l/c}
\right)
r_{21} e^{i n_2 \omega | x - (-l)/2 | /c}
\right]
\right.
\nonumber
\\ & &
\left.
+\Theta \left( x - {\textstyle{\frac{1}{2}}} l \right) \vartheta
\left(
e^{i n_2 \omega | x' - (-l)/2 | /c} r_{21}
e^{i n_2 \omega l/c}
+ e^{i n_2 \omega | l/2 - x' | /c}
\right)
t_{23} e^{i n_3 \omega | x - l/2 | /c}
\right\}
\nonumber
\\ & &
+ \left[ 2 i n_3 \frac{\omega}{c} \right]^{-1}
\Theta \left(x' - {\textstyle{\frac{1}{2}}} l \right)
\left\{
\Theta \left( - x - {\textstyle{\frac{1}{2}}} l \right)
e^{i n_3 \omega | x' - l/2 | /c} t_{32}
\vartheta e^{i n_2 \omega l/c} t_{21}
e^{i n_1 \omega | (-l)/2 - x | /c}
\right.
\nonumber
\\ & &
\left.
+ \left[
\Theta \left( x + {\textstyle{\frac{1}{2}}} l \right)
- \Theta \left( x - {\textstyle{\frac{1}{2}}} l \right)
\right]
e^{i n_3 \omega | x' - l/2 |/c} t_{32}
\vartheta
\left(
e^{i n_2 \omega | l/2 - x | /c}
\right.
\right.
\nonumber
\\ & &
\left.
\left.
+ e^{i n_2 \omega l/c} r_{21}
e^{i n_2 \omega | x - (-l)/2 | /c}
\right)
+ \Theta \left( x - {\textstyle{\frac{1}{2}}} l \right)
\left[
e^{i n_3 \omega | x - x'|/c }
\right.
\right.
\nonumber
\\ & &
\left.
\left.
+ e^{i n_3 \omega | x' - l/2 | /c }
\left(
r_{32} + t_{32} \vartheta e^{i n_2 \omega l/c}
r_{21} e^{i n_2 \omega l/c} t_{23}
\right)
e^{i n_3 \omega | x - l/2 | /c }
\right]
\right\}.
\label{a.4}
\end{eqnarray}
Here, the interface reflection and transmission coefficients
$r_{ij}$ $\!\equiv$ $\!r_{ij}(\omega)$ and $t_{ij}$ $\!\equiv$
$\!t_{ij}(\omega)$, respectively, are defined by
\begin{equation}
r_{ij} (\omega) = - r_{ji} (\omega)
= \frac{n_i (\omega) - n_j (\omega)}{n_i (\omega) + n_j (\omega)}\,,
\label{a.1}
\end{equation}
\begin{equation}
t_{ij} (\omega) =  \frac{2 n_i (\omega) }{n_i (\omega) + n_j (\omega)}\,,
\label{a.2}
\end{equation}
and the factor $\vartheta\!\equiv\!\vartheta (\omega)$, which arises
from multiple reflections inside the slab, reads as
\begin{eqnarray}
\lefteqn{
\vartheta (\omega) = \sum_{j=0}^{\infty}
\left[
e^{i n_2 (\omega) \omega l/c}  r_{21} (\omega)
e^{i n_2 (\omega) \omega l/c} r_{23} (\omega) \right]^j
}
\nonumber
\\ & &
\hspace{5.5ex}
= \left[
1 - e^{i n_2 (\omega) \omega l/c} r_{21} (\omega)
e^{i n_2 (\omega) \omega l/c} r_{23} (\omega) \right]^{-1}
\hspace{-2.5ex}.
\label{a.3}
\end{eqnarray}
Note that the above given form of the Green function permits of
a direct physical interpretation. The terms in Eq.~(\ref{a.4}) simply
correspond to the potential propagations of radiation from a source
point $x'$ to a point of observation, $x$.

\section{Photonic operators, noise oper\-ators, in\-put--ou\-tput
relations}
\subsection{Single-slab dielectric plate}
\label{expl}
\setcounter{equation}{0}
Substituting in Eq.~(\ref{2.8}) for $G(x,x',\omega)$ the expression
(\ref{a.4}) and rewriting the result (within the space
intervals $-\infty\!\leq\!x\!\leq\!-l/2$ and $l/2\!\leq\!x\!\leq\!\infty$)
in the form (\ref{3.7}), we easily see that the amplitude operators of the
incoming fields from the left and right, respectively,
$\hat{a}_{1+}(x,\omega)$ and $\hat{a}_{3-}(x,\omega)$, read as
\begin{equation}
\hat{a}_{1+}(x, \omega)
= \frac{1}{i} \sqrt{2 \gamma_1 (\omega) \frac{\omega}{c}} \,
e^{-\gamma_1 (\omega) \omega x/c}
\int_{-\infty}^{x} {\rm d} x' \, e^{- i n_1 (\omega) \omega x'/c}
\hat{f} (x',\omega),
\label{3.10a}
\end{equation}
\begin{equation}
\hat{a}_{3-}(x, \omega)
= \frac{1}{i} \sqrt{2 \gamma_3 (\omega) \frac{\omega}{c}} \,
e^{\gamma_3 (\omega) \omega x/c}
\int_{x}^{\infty} {\rm d} x' \, e^{i n_3 (\omega) \omega x'/c}
\hat{f} (x',\omega)
\label{3.10b}
\end{equation}
$[n_j(\omega)\!=\!\beta_j(\omega)\!+\!i\gamma_j(\omega)]$,
and the amplitude operators of the outgoing fields
to the left and right, respectively,
$\hat{a}_{1-}(x,\omega)$ and $\hat{a}_{3+}(x,\omega)$,
are given by
\begin{eqnarray}
\lefteqn{
\hat{a}_{1-} (x, \omega) = \frac{1}{i} \sqrt{2 \gamma_1 (\omega)
\frac{\omega}{c}} \,
e^{\gamma_1 (\omega) \omega x/c}
\int_{x}^{\frac{l}{2}} {\rm d} x' \, e^{i n_1 (\omega) \omega x'/c}
\hat{f} (x', \omega)
}
\nonumber
\\ & &
\hspace{2ex}
+e^{\gamma_1 (\omega) \omega( x - l/2 ) / c}
e^{-i n_1 (\omega) \omega l/c}
\left[
r_{12} (\omega) + t_{12} (\omega) r_{23} (\omega) t_{21} (\omega)
e^{2 i n_2 (\omega) \omega l/c} \vartheta (\omega)
\right]
\hat{a}_{1+}\! \left( -\textstyle{\frac{1}{2}} l, \omega \right)
\nonumber
\\ & &
\hspace{2ex}
+ \sqrt{2 \gamma_1 (\omega)} \,
e^{\gamma_1 (\omega) \omega x/c} \,
\frac{n_1 (\omega)}{n_2 (\omega)}
\sqrt{\frac{\beta_{2}(\omega)\gamma_{2}(\omega)}
{\beta_{1}(\omega)\gamma_{1}(\omega)}} \,
\vartheta (\omega) t_{21} (\omega)
e^{- i n_1 (\omega) \omega l/(2c)}
\nonumber
\\ & &
\hspace{40ex}
\times \,
\left[ \hat{g}'_{-}(\omega) + r_{23}
(\omega) e^{i n_2 (\omega) \omega l/c} \, \hat{g}'_{+}(\omega)
\right]
\nonumber
\\ & &
\hspace{2ex}
+e^{ [ \gamma_1 (\omega) x - \gamma_3 (\omega) l/2 ] \omega/c } \,
\frac{n_1 (\omega)}{n_3 (\omega)}
\sqrt{\frac{\beta_3 (\omega)}{\beta_1 (\omega)}} \,
\nonumber
\\ & &
\hspace{20ex}
\times \,
e^{-i [ n_1 (\omega) - 2 n_2 (\omega) + n_3 (\omega) ]
     \omega l/(2c)} t_{32} (\omega) t_{21} (\omega)
\vartheta (\omega) \,
\hat{a}_{3-}\!\left( \textstyle{\frac{1}{2}}l, \omega \right)\!,
\label{3.10d}
\end{eqnarray}
\begin{eqnarray}
\lefteqn{
\hat{a}_{3+} (x, \omega) = \frac{1}{i} \sqrt{2 \gamma_3 (\omega)
\frac{\omega}{c}} \,
e^{- \gamma_3 (\omega) \omega x/c}
\int_{\frac{l}{2}}^{x} {\rm d} x' \, e^{-i n_3 (\omega) \omega x'/c}
\hat{f} (x', \omega)
}
\nonumber
\\ & &
\hspace{2ex}
+e^{-\gamma_3 (\omega) \omega ( x + l/2 ) /c}
e^{-i n_3 (\omega) \omega l/c}
\left[
r_{32} (\omega) + t_{32} (\omega) r_{21} (\omega) t_{23} (\omega)
e^{2 i n_2 (\omega) \omega l/c} \vartheta (\omega)
\right]
\hat{a}_{3-}\!\left( \textstyle{\frac{1}{2}} l, \omega \right)
\nonumber
\\ & &
\hspace{2ex}
+ \sqrt{2 \gamma_3 (\omega)} \,
e^{-\gamma_3 (\omega) \omega x/c} \,
\frac{n_3 (\omega)}{n_2 (\omega)}
\sqrt{\frac{\beta_{2}(\omega)\gamma_{2}(\omega)}
{\beta_{3}(\omega)\gamma_{3}(\omega)}} \,
\vartheta (\omega) t_{23} (\omega)
e^{- i n_3 (\omega) \omega l/(2c)}
\nonumber
\\ & &
\hspace{40ex}
\times \,
\left[ \hat{g}'_{+}(\omega) + r_{21}
(\omega) e^{i n_2 (\omega) \omega l/c} \, \hat{g}'_{-}(\omega)
\right]
\nonumber
\\ & &
\hspace{2ex}
+e^{-[ \gamma_1 (\omega) l/2 + \gamma_3 (\omega) x ] \omega/c } \,
\frac{n_3 (\omega)}{n_1 (\omega)}
\sqrt{\frac{\beta_1 (\omega)}{\beta_3 (\omega)}} \,
\nonumber \\
& &
\hspace{20ex}
\times \,
e^{-i [ n_1 (\omega) - 2 n_2 (\omega) + n_3 (\omega) ] \omega l/(2c)}
t_{12} (\omega) t_{23} (\omega) \vartheta (\omega) \,
\hat{a}_{1+}\!\left( -\textstyle{\frac{1}{2}} l, \omega \right)\!,
\label{3.10e}
\end{eqnarray}
where the operators $\hat{g}'_\pm(\omega)$ are defined
in Eq.~(\ref{3.10}).

Inverting the relations (\ref{3.40}), we may express the operators
$\hat{g}'_\pm(\omega)$ in terms of the operators
$\hat{g}_\pm^{(1)}(\omega)$ and rewrite Eqs.~(\ref{3.10d})
(for $x\!=\!-l/2$) and (\ref{3.10e}) (for $x\!=\!l/2$) in the
compact form
\begin{equation}
\left(\!\!\!
\begin{array}{c}
\hat{a}_{1-}\big(-\textstyle{\frac{1}{2}}l,\omega\big)
\\[.5ex]
\hat{a}_{3+}\big(\textstyle{\frac{1}{2}}l,\omega\big)
\end{array}
\!\!\!\right) \! = \! \widetilde{\bf T}^{(1)} \!\left(\!\!\!
\begin{array}{c}
\hat{a}_{1+}\big(-\textstyle{\frac{1}{2}}l,\omega\big)
\\[.5ex]
\hat{a}_{3-}\big(\textstyle{\frac{1}{2}}l,\omega\big)
\end{array}
\!\!\!\right) \! + \! \widetilde{\bf A}^{(1)}\! \left(\!\!
\begin{array}{c}
\hat{g}_{+}^{(1)}(\omega)
\\[.5ex]
\hat{g}_{-}^{(1)}(\omega)
\end{array}
\!\!\right)\!.
\label{a.4c}
\end{equation}
The elements of the characteristic transformation matrix
$\widetilde{\bf T}^{(1)}$, $T_{ik}^{(1)}(\omega)$, are seen to be
\begin{eqnarray}
\lefteqn{
T_{11}^{(1)}(\omega) = e^{-i \beta_1 (\omega) \omega l/c}
}
\nonumber
\\ & &
\hspace{1ex}
\times  \left[ r_{12}(\omega) + t_{12}(\omega)
e^{2 i n_2 \omega \omega l/c}
r_{23} (\omega) \vartheta(\omega) t_{21} (\omega) \right]\!,
\label{a.5a}
\\ & &
T_{12}^{(1)}(\omega) = \frac{n_1 (\omega)}{n_3 (\omega)}
\sqrt{\frac{\beta_3 (\omega)}{\beta_1 (\omega)}} \,
e^{- i [ \beta_1 (\omega) + \beta_3 (\omega) ] \omega l/(2c)}
\nonumber
\\ & &
\hspace{1ex}
\times \, t_{32} (\omega)
e^{i n_2 (\omega) \omega l/c}
\vartheta(\omega) t_{21} (\omega) ,
\label{a.6a}
\\ & &
T_{21}^{(1)}(\omega) = \frac{n_3(\omega)}{n_1(\omega)}
\sqrt{\frac{\beta_1 (\omega)}{\beta_3(\omega)}} \,
e^{- i [ \beta_1 (\omega) + \beta_3 (\omega) ] \omega l/(2c)}
\nonumber
\\ & &
\hspace{1ex}
\times \, t_{12} (\omega)
e^{i n_2 (\omega) \omega l/c}
\vartheta(\omega) t_{23} (\omega),
\label{a.7a}
\\ & &
T_{22}^{(1)}(\omega) = e^{-i \beta_3 (\omega) \omega l/c}
\nonumber
\\ & &
\hspace{1ex}
\times  \left[ r_{32} (\omega) + t_{32} (\omega)
e^{2 i n_2 (\omega) \omega l/c} r_{21} (\omega)
\vartheta(\omega) t_{23} (\omega) \right]\!,
\label{a.8a}
\end{eqnarray}
and the elements of the characteristic absorption matrix
$\widetilde{\bf A}^{(1)}$, $A_{ik}^{(1)}(\omega)$, read as
\begin{eqnarray}
\lefteqn{
A_{11}^{(1)}(\omega) =
\sqrt{\gamma_2(\omega)\frac{\beta_2(\omega)}{\beta_1(\omega)}} \,
e^{-i \beta_1 (\omega) \omega l/(2c)}
}
\nonumber
\\ & &
\hspace{4ex}
\times \,
t_{12}(\omega) \vartheta(\omega) \sqrt{c_{+}}
\left[ 1 + e^{i n_2 (\omega) \omega l/c}
r_{23} (\omega) \right]\!,
\label{a.9a}
\\ & &
A_{12}^{(1)}(\omega) =
\sqrt{\gamma_2(\omega)\frac{\beta_2(\omega)}{\beta_1(\omega)}} \,
e^{-i \beta_1 (\omega) \omega l/(2c)}
\nonumber
\\ & &
\hspace{4ex}
\times \,
t_{12} (\omega)  \vartheta(\omega) \sqrt{c_{-}}
\left[ 1 - e^{i n_2 (\omega) \omega l/c}
r_{23} (\omega) \right]\!,
\label{a.10a}
\\ & &
A_{21}^{(1)}(\omega) =
\sqrt{\gamma_2(\omega)\frac{\beta_2(\omega)}{\beta_3(\omega)}} \,
e^{-i \beta_3 (\omega) \omega l/(2c)}
\nonumber
\\ & &
\hspace{4ex}
\times \,
t_{32} (\omega) \vartheta(\omega) \sqrt{c_{+}}
\left[ e^{i n_2 (\omega) \omega l/c}
r_{21} (\omega) + 1 \right]\!,
\label{a.11a}
\\ & &
A_{22}^{(1)}(\omega) =
\sqrt{\gamma_2(\omega)\frac{\beta_2(\omega)}{\beta_3(\omega)}} \,
e^{-i \beta_3 (\omega) \omega l/(2c)}
\nonumber
\\ & &
\hspace{4ex}
\times \,
t_{32} (\omega) \vartheta(\omega) \sqrt{c_{-}}
\left[ e^{i n_2 (\omega) \omega l/c}
r_{21} (\omega) - 1\right]\!.
\label{a.12a}
\end{eqnarray}
Note that when the plate is surrounded by vacuum,
\begin{equation}
  n_1(\omega) = n_3(\omega) \equiv 1,
\label{a.3a}
\end{equation}
so that
\begin{equation}
  \beta_1(\omega) = \beta_3(\omega) \equiv 1,  \quad
  \gamma_1(\omega) = \gamma_3(\omega) \equiv 0,
\label{a.3b}
\end{equation}
the following relations are valid [cf. Eqs.~(\ref{a.1}) -- (\ref{a.3})]:
\begin{eqnarray}
& &
r_{12} (\omega) \! =\! r_{32} (\omega) \! = \!
\frac{1 - n_2 (\omega)}{1 + n_2 (\omega)}
\! = \! - r_{21} (\omega) \! = \! - r_{23} (\omega),
\label{b.1}
\\
\lefteqn{
t_{12} (\omega) = t_{32} (\omega) = \frac{2}{1 + n_2 (\omega)}\,,
}
\label{b.2}
\\ & &
t_{21} (\omega) = t_{23} (\omega)
= \frac{2 n_2 (\omega)}{1 + n_2 (\omega)}\,,
\label{b.3}
\\ & &
\vartheta (\omega) = \sum_{j=0}^{\infty}
\left[
e^{i n_2 (\omega) \omega l/c} r_{21} (\omega)
e^{i n_2 (\omega) \omega l/c} r_{23} (\omega) \right]^j
\nonumber
\\ & &
\hspace{5.5ex}
= \left[1 - r_{21}^2 (\omega)
e^{2 i n_2 (\omega) \omega l/c} \right]^{-1}
\hspace{-2.5ex}.
\label{b.4}
\end{eqnarray}

Using Eqs.~(\ref{3.40}) and (\ref{3.10}) and recalling the
commutation relations (\ref{2.16}) and (\ref{2.17}),
from Eqs.~(\ref{3.10a}) and  (\ref{3.10b}) we derive that
\begin{equation}
\big[ \hat{a}_{1+}(x, \omega), \hat{a}_{1+}^{\dagger} (x', \omega')
\big]  =  e^{-\gamma_1(\omega)|x-x'|/c} \delta(\omega-\omega'),
\label{a.4d}
\end{equation}
\begin{equation}
\big[ \hat{a}_{3-}(x, \omega), \hat{a}_{3-}^{\dagger} (x', \omega')
\big]  =  e^{-\gamma_3(\omega)|x-x'|/c} \delta(\omega-\omega'),
\label{a.4e}
\end{equation}
\begin{equation}
\big[ \hat{a}_{1+}(x, \omega), \hat{a}_{3-}^{\dagger} (x', \omega')
\big]  =  0,
\label{a.4a}
\end{equation}
\begin{equation}
\left[ \hat{a}_{1+}(x, \omega),
   \big( \hat{g}_{\pm}^{(1)} (\omega') \big)^{\dagger}
\right] = 0 = \left[ \hat{a}_{3-}(x', \omega),
   \big( \hat{g}_{\pm}^{(1)} (\omega') \big)^{\dagger}
\right].
\label{a.4b}
\end{equation}
The commutation relations for the output amplitude operators
$\hat{a}_{1-}(x,\omega)$, $\hat{a}_{3+}(x,\omega)$ and
$\hat{a}_{1-}^\dagger(x,\omega)$, $\hat{a}_{3+}^\dagger(x,\omega)$
can be derived in a similar way or, more easily, using Eqs.~(\ref{3.7a})
and (\ref{3.7b}) and applying the input--output relations (\ref{a.4c})
(together with the corresponding commutation relations).
Straightforward calculation yields
\begin{eqnarray}
\lefteqn{
\big[ \hat{a}_{1-} ( x,\omega ) ,
\hat{a}^{\dagger}_{1-} ( x',\omega' ) \big] =
\delta( \omega-\omega' )
\bigg\{   e^{-\gamma_1( \omega ) \omega | x-x' | /c }
}
\nonumber
\\ & &
+e^{-\gamma_1( \omega ) \omega ( | x-( -l/2 )|/c
+ | x'-( -l/2) | / c )}
\bigg[
\big| T^{(1)}_{11}(\omega) \big|^2
+\big| T^{(1)}_{12}(\omega) \big|^2 + \big| A^{(1)}_{11}(\omega) \big|^2
+\big| A^{(1)}_{12}(\omega) \big|^2 - 1
\nonumber
\\ & &
+i \frac{\gamma_1(\omega)}{\beta_1(\omega)}
\left[  T^{(1)}_{11}(\omega)
\big( e^{i\beta_1 (\omega) \omega l}
- e^{-2i\beta_1 (\omega) \omega x'} \big)
- \big(T_{11}^{(1)} (\omega)\big)^\ast
\big( e^{-i\beta_1 (\omega) \omega l}
- e^{2i\beta_1 (\omega) \omega x} \big) \right]
\bigg]
\bigg\}\!,
\nonumber \\ &&
\label{3.17}
\end{eqnarray}
\begin{eqnarray}
\lefteqn{
\big[ \hat{a}_{3+} ( x,\omega ) ,
\hat{a}^{\dagger}_{3+} ( x',\omega' ) \big] =
\delta( \omega-\omega' )
\bigg\{   e^{-\gamma_3( \omega ) \omega | x-x' | /c }
}
\nonumber
\\ & &
+e^{-\gamma_3( \omega ) \omega ( | x-l/2 |/c
+ | x' - l/2 | / c )}
\bigg[
\big| T^{(1)}_{21}(\omega) \big|^2
+\big| T^{(1)}_{22}(\omega) \big|^2 + \big| A^{(1)}_{21}(\omega) \big|^2
+\big| A^{(1)}_{22}(\omega) \big|^2 - 1
\nonumber
\\ & &
+i \frac{\gamma_3(\omega)}{\beta_3(\omega)}
\left[ T^{(1)}_{22} (\omega)
\big( e^{i\beta_3 (\omega) \omega l}
- e^{2i\beta_3 (\omega) \omega x'} \big)
- \big(T_{22}^{(1)} (\omega)\big)^\ast
\big( e^{-i\beta_3 (\omega) \omega l}
- e^{-2i\beta_3 (\omega) \omega x} \big) \right]
\bigg]
\bigg\}\!,
\nonumber \\ &&
\label{3.17b}
\end{eqnarray}
\begin{eqnarray}
\lefteqn{
\big[ \hat{a}_{3+} ( x,\omega ) ,
\hat{a}^{\dagger}_{1-} ( x',\omega' ) \big] =
\delta(\omega - \omega')
\bigg\{
e^{- \gamma_1 (\omega) \omega | x' - (-l/2) |/c
- \gamma_3 (\omega) \omega | x - l/2 |/c )  }
}
\nonumber
\\ &  &
\times \,
\bigg[
\big(T^{(1)}_{11} (\omega)\big)^\ast T^{(1)}_{21} (\omega)
+ \big(T^{(1)}_{12} (\omega)\big)^\ast T^{(1)}_{22} (\omega)
+ \big(A^{(1)}_{11} (\omega)\big)^\ast A^{(1)}_{21} (\omega)
+ \big(A^{(1)}_{12} (\omega)\big)^\ast A^{(1)}_{22} (\omega)
\nonumber
\\ & &
\hspace{4ex}
+ \, i \, T^{(1)}_{21} (\omega)
\, \frac{\gamma_1 (\omega)}{\beta_1 (\omega)}
\left(
e^{i\beta_1(\omega) \omega l/c} - e^{-2i \beta_1 (\omega) \omega x'/c}
\right)
\nonumber
\\ & &
\hspace{26ex}
 + \, i \, \big(T^{(1)}_{12} (\omega) \big)^\ast
\frac{\gamma_3 (\omega)}{\beta_3 (\omega)}
\left(
e^{-2i\beta_3(\omega) \omega x/c} - e^{-i \beta_3
(\omega) \omega l/c}
\right)
\bigg]
\bigg\}\!.
\nonumber \\ &&
\label{3.18}
\end{eqnarray}
Needless to say that $\hat{a}_{1-}(x,\omega)$ and $\hat{a}_{3+}(x,\omega)$
are commuting quantities. Using Eqs.~(\ref{a.5a}) -- (\ref{a.12a})
and assuming that the plate is embedded in non-absorbing media
($\gamma_1(\omega)\!=\!\gamma_3(\omega)\!=\!0$), the following
relations are easily proved correct:
\begin{eqnarray}
\lefteqn{
\big| T_{11}^{(1)} (\omega) \big|^2
+ \big| T_{12}^{(1)} (\omega) \big|^2
+ \big| A_{11}^{(1)} (\omega) \big|^2
+ \big| A_{12}^{(1)} (\omega) \big|^2
}
\nonumber
\\ & &
=\big| T_{21}^{(1)} (\omega) \big|^2
+ \big| T_{22}^{(1)} (\omega) \big|^2
+ \big| A_{21}^{(1)} (\omega) \big|^2
+ \big| A_{22}^{(1)} (\omega) \big|^2 = 1,
\nonumber \\ &&
\label{3.18a}
\end{eqnarray}
\begin{eqnarray}
\lefteqn{
T_{11}^{(1)} (\omega) \big( T_{21}^{(1)} (\omega) \big)^*
+ T_{12}^{(1)} (\omega) \big( T_{22}^{(1)} (\omega) \big)^*
}
\nonumber
\\ & &
\hspace{2ex}
+ A_{11}^{(1)} (\omega) \big( A_{21}^{(1)} (\omega) \big)^*
+ A_{12}^{(1)} (\omega) \big( A_{22}^{(1)} (\omega) \big)^* = 0.
\label{3.18b}
\end{eqnarray}
In this case, both the input-field commutation relations
(\ref{a.4d}) -- (\ref{a.4a}) and the output-field commutation
relations (\ref{3.17}) -- (\ref{3.18}) obviously reduce to the
familiar bosonic commutation relations for photon destruction and
creation operators.

\subsection{Multi-slab dielectric plates}
\label{induc}
Starting from the results derived in
\ref{expl} for a single-slab
dielectric plate ($N\!=\!3$), the results for an arbitrary dielectric
plate ($N\!\geq\!3$, cf.
Fig.~2)
can be obtained step by step,
without explicitly calculating the multi-slab Green function.
For this purpose we show that when Eqs.~(\ref{3.27}) -- (\ref{3.27k})
are valid for $N\!-\!1$, then they are also valid for $N$.
Using Eq.~(\ref{3.27}), with $N\!-\!1$ in place of $N$, and expressing
the operators $ \hat{a}_{N-1\,\pm}(x_{N-2},\omega) $ in terms of the
operators $ \hat{a}_{1\pm}(x_1,\omega) $, we find that
\begin{equation}
\left(
\begin{array}{c}
\hat{a}_{N-1\,+}(x_{N-2},\omega)
\\[.5ex]
\hat{a}_{N-1\,-}(x_{N-2},\omega)
\end{array}
\right)
= \widetilde{\bf P} \left(
\begin{array}{c}
\hat{a}_{1-}(x_1,\omega)
\\[.5ex]
\hat{a}_{1+}(x_1,\omega)
\end{array}
\right)
+ \widetilde{\bf Q} \left(
\begin{array}{c}
\hat{g}^{(N-3)}_{+}(\omega)
\\[.5ex]
\hat{g}^{(N-3)}_{-}(\omega)
\end{array}
\right)\!,
\label{3.53}
\end{equation}
where the matrices $ \widetilde{\bf P}\!\equiv\!
\widetilde{\bf P}^{(N-3)} $ and $ \widetilde{\bf Q}\!\equiv\!
\widetilde{\bf Q}^{(N-3)} $ are given by
\begin{eqnarray}
\widetilde{\bf P} & = &
\left(
T_{12}^{(N-3)}
\right)^{-2} \left(\!
\begin{array}{c@{\hspace{4ex}}c}
T_{22}^{(N-3)} &  T_{21}^{(N-3)} T_{12}^{(N-3)} \!
- \! T_{22}^{(N-3)} T_{11}^{(N-3)}
\\[.5ex]
1 & - T_{11}^{(N-3)}
\end{array}
\!\right)\!,
\label{3.54}
\end{eqnarray}
\begin{eqnarray}
\lefteqn{
\widetilde{\bf Q}  =
\left(
T_{12}^{(N-3)}
\right)^{-2}
}
\nonumber \\
&&
\times \
\left(\!
\begin{array}{c@{\hspace{4ex}}c}
A_{21}^{(N-3)} T_{12}^{(N-3)} \! - \! A_{11}^{(N-3)} T_{22}^{(N-3)}
& A_{22}^{(N-3)} T_{12}^{(N-3)} \! - \! A_{12}^{(N-3)} T_{22}^{(N-3)}
\\[.5ex]
- A_{11}^{(N-3)} & - A_{12}^{(N-3)}
\end{array}
\!\right)\!.
\hspace{5ex}
\label{3.55}
\end{eqnarray}

The desired $(N\!-\!2)$-slab plate can be obtained by
supplement to the $(N\!-\!3)$-slab plate of the $(N\!-\!1)$th slab
(cf. Fig.~2).
Applying Eq.~(\ref{3.7a}) to
$\hat{a}_{N-1\,\pm}(x,\omega)$ ($x_{N-2}\!\leq\!x\!\leq\!x_{N-1}$) yields
\begin{equation}
\hat{a}_{N-1\,\pm} (x_{N-1},\omega)
= \hat{a}_{N-1\,\pm} (x_{N-2},\omega)
e^{\mp \gamma_{N-1}\!(\omega)\, \omega l_{N-1} / c}
+ \hat{d}'_{N-1\,\pm}(\omega),
\label{3.55a}
\end{equation}
where $ l_{N-1} \! = \! x_{N-1} \! - \! x_{N-2} $, and
\begin{equation}
\hat{d}'_{N-1\,\pm}(\omega)
= \int_{x_{N-2}}^{x_{N-1}} {\rm d}y \,
\hat{F}_{N-1\,\pm} (y, \omega)
e^{\mp \gamma_{N-1}\!(\omega) \, \omega \left( x_{N-1} - y \right) / c },
\label{3.56}
\end{equation}
with $\hat{F}_{N-1\,\pm} (y, \omega)$ according to Eq.~(\ref{3.7b}).
Recalling the commutation relations (\ref{2.25}), the operators
$\hat{d}'_{N-1\,\pm}$ are found to satisfy the commutation relations
\begin{equation}
\left[
\hat{d}'_{N-1\,\pm}(\omega),
\hat{d}'\,\!_{\!\! N-1\,\pm}^{\dagger} (\omega')
\right] = 2 \delta(\omega - \omega')
e^{\mp \gamma_{N-1}\!(\omega)\, \omega l_{N-1} / c }
\sinh{\!\left[
\gamma_{N-1} (\omega) \frac{\omega}{c} l_{N-1}
\right]},
\label{3.57}
\end{equation}
\begin{eqnarray}
\lefteqn{
\left[
\hat{d}'_{N-1\,\pm}(\omega),
\hat{d}'\,\!_{\!\! N-1\,\mp}^{\dagger} (\omega')
\right] =  - 2 \delta(\omega - \omega')
 \frac{\gamma_{N-1} (\omega)}{\beta_{N-1}
(\omega)}
}
\nonumber
\\ & &
\hspace{7ex} \times \,
 e^{\mp i \beta_{N-1}\!(\omega)\,\omega
\left( x_{N-2} + x_{N-1} \right) / c} \sin{\!\left[
\beta_{N-1} (\omega) \frac{\omega}{c} l_{N-1}
\right]}.
\label{3.58}
\end{eqnarray}
By means of linear combination of the operators $\hat{d}'_{N-1\,\pm}$,
bosonic operators $\hat{d}_{N-1\,\pm}$ can be introduced as
\begin{equation}
\left(
\begin{array}{c}
\hat{d}_{N-1\,+}(\omega)
\\[.5ex]
\hat{d}_{N-1\,-}(\omega)
\end{array}
\right) = \widetilde{\bf D} \left(
\begin{array}{c}
\hat{d}'_{N-1\,+}(\omega)
\\[.5ex]
\hat{d}'_{N-1\,-}(\omega)
\end{array}
\right)\!,
\label{3.63}
\end{equation}
\begin{eqnarray}
\left[
\hat{d}_{N-1\,\pm} (\omega), \hat{d}_{N-1\,\pm}^{\dagger} (\omega')
\right]  & = & \delta( \omega - \omega' ) ,
\label{3.65}
\\
\left[
\hat{d}_{N-1\,\pm} (\omega), \hat{d}_{N-1\,\mp}^{\dagger} (\omega')
\right] & = & 0.
\label{3.66}
\end{eqnarray}
In Eq.~(\ref{3.63}), the matrix $\widetilde{\bf D}$ reads as
\begin{eqnarray}
\lefteqn{
D_{11} =
\left\{
2 \left[ \alpha_{+} + \left(
\frac{\alpha_{+}}{\alpha_{-}} \right)^{\!\frac{1}{2}}
\left| \alpha_{0} \right|
\right]
\right\}^{-\frac{1}{2}}\hspace{-3ex},
}
\label{3.61}
\\ & &
D_{21} =
-\left\{
2 \left[ \alpha_{+} - \left(
\frac{\alpha_{+}}{\alpha_{-}} \right)^{\!\frac{1}{2}}
\left| \alpha_{0} \right|
\right]
\right\}^{-\frac{1}{2}}\hspace{-3ex},
\label{3.61a}
\\ & &
D_{12} =
 \exp{\left[i \arg{\!\left[
\alpha_{0}
\right]}\right]}
\left\{
2 \left[ \alpha_{-} + \left(
\frac{\alpha_{-}}{\alpha_{+}} \right)^{\!\frac{1}{2}}
\!\left| \alpha_{0} \right|
\right]
\right\}^{\!-\frac{1}{2}}\hspace{-3ex},
\label{3.62}
\\ & &
D_{22} =
\exp{\left[i \arg{\!\left[
\alpha_{0}
\right]}\right]} \left\{
2 \left[ \alpha_{-} - \left(
\frac{\alpha_{-}}{\alpha_{+}} \right)^{\!\frac{1}{2}}
\!\left| \alpha_{0} \right|
\right]
\right\}^{\!-\frac{1}{2}}\hspace{-3ex},
\label{3.62a}
\end{eqnarray}
where
\begin{equation}
{\alpha}_{\pm}
=  2 e^{\mp \gamma_{N-1}\!(\omega)\,\omega l_{N-1} / c}
\sinh{\!\left[
\gamma_{N-1} (\omega) \frac{\omega}{c} l_{N-1}
\right]},
\label{3.59}
\end{equation}
\begin{equation}
{\alpha}_{0}
= - 2 \frac{\gamma_{N-1}
(\omega)}{\beta_{N-1}
(\omega)} e^{- i \beta_{N-1}\!(\omega)\,\omega \left( x_{N-2} + x_{N-1} \right)
 / c }
\sin{\!\left[
\beta_{N-1} (\omega) \frac{\omega}{c} l_{N-1}
\right]}.
\label{3.60}
\end{equation}
Combining Eqs.~(\ref{3.55a}) and (\ref{3.63}) we may write
\begin{equation}
\left(
\begin{array}{c}
\hat{a}_{N-1\,+}(x_{N-1},\omega)
\\[.5ex]
\hat{a}_{N-1\,-}(x_{N-1},\omega)
\end{array}
\right) = \widetilde{\bf R} \left(
\begin{array}{c}
\hat{a}_{N-1\,+}(x_{N-2},\omega)
\\[.5ex]
\hat{a}_{N-1\,-}(x_{N-2},\omega)
\end{array}
\right)
+ \widetilde{\bf D}^{-1} \left(
\begin{array}{c}
\hat{d}_{N-1\,+}(\omega)
\\
\hat{d}_{N-1\,-}(\omega)
\end{array}
\right)\!,
\label{3.67}
\end{equation}
where $R_{ii'}\!=\!R_{ii}\,\delta_{ii'}$, with
\begin{equation}
R_{11} = R_{22}^{-1} = \exp{\left[
- \gamma_{N-1} (\omega) \frac{\omega}{c} l_{N-1}
\right]}.
\end{equation}

We now relate the operators $ \hat{a}_{N\,\pm} (x_{N-1}, \omega) $
and $ \hat{a}_{N-1\,\pm} (x_{N-1},\omega)$ to each other, applying
Eqs.~(\ref{3.7}) and (\ref{3.7a}) and recalling that $\hat{A}(x)$ is
continuously differentiable at $ x_{N-1}$. Straightforward calculation
yields
\begin{equation}
\left(
\begin{array}{c}
\hat{a}_{N+}(x_{N-1},\omega)
\\[.5ex]
\hat{a}_{N-}(x_{N-1},\omega)
\end{array}
\right) = \widetilde{\bf S} \left(
\begin{array}{c}
\hat{a}_{N-1\,+}(x_{N-1},\omega)
\\[.5ex]
\hat{a}_{N-1\,-}(x_{N-1},\omega)
\end{array}
\right)\!,
\label{3.69}
\end{equation}
where the elements of the matrix $ \widetilde{\bf S} $ read as
\begin{eqnarray}
S_{11} & = &
\sqrt{\frac{\beta_{N-1}(\omega)}{
\beta_{N} (\omega)}}
\frac{n_N (\omega) + n_{N-1} (\omega)}{2 n_{N-1} (\omega)}\,
e^{- i \left[
\beta_{N}(\omega) - \beta_{N-1}\!(\omega)
\right] \omega x_{N-1} / c},
\label{3.70}
\\
S_{12} & = &
\sqrt{\frac{\beta_{N-1}(\omega)}{
\beta_{N} (\omega)}}
\frac{n_N (\omega) - n_{N-1} (\omega)}{2 n_{N-1} (\omega)}\,
e^{- i \left[
\beta_{N}(\omega) + \beta_{N-1}\!(\omega)
\right] \omega x_{N-1} / c},
\label{3.71}
\\
S_{21} & = &
\sqrt{\frac{\beta_{N-1} (\omega)}{
\beta_{N} (\omega)}}
\frac{n_N (\omega) - n_{N-1} (\omega)}{2 n_{N-1} (\omega)} \,
e^{ i \left[
\beta_{N}(\omega) + \beta_{N-1}\!(\omega)
\right] \omega x_{N-1} / c},
\label{3.72}
\\
S_{22} & = &
\sqrt{\frac{\beta_{N-1} (\omega)}{
\beta_{N} (\omega)}}
\frac{n_N (\omega) + n_{N-1} (\omega)}{2 n_{N-1} (\omega)} \,
e^{ i \left[
\beta_{N}(\omega) - \beta_{N-1}\!(\omega)
\right] \omega x_{N-1} / c}.
\label{3.73}
\end{eqnarray}
Combining Eqs.~(\ref{3.53}), (\ref{3.67}), and (\ref{3.69}), we obtain
\begin{eqnarray}
\lefteqn{
\left(
\begin{array}{c}
\hat{a}_{N+}(x_{N-1},\omega)
\\[.5ex]
\hat{a}_{N-}(x_{N-1},\omega)
\end{array}
\right) = \widetilde{\bf S} \widetilde{\bf R}
\widetilde{\bf P} \left(
\begin{array}{c}
\hat{a}_{1-}(x_1,\omega)
\\[.5ex]
\hat{a}_{1+}(x_1,\omega)
\end{array}
\right)
}
\nonumber
\\[.5ex]
& &
\hspace{1ex}
+ \, \widetilde{\bf S} \widetilde{\bf R}
\widetilde{\bf Q} \left(
\begin{array}{c}
\hat{g}^{(N-3)}_{+}(\omega)
\\[.5ex]
\hat{g}^{(N-3)}_{-}(\omega)
\end{array}
\right)
+ \widetilde{\bf S}  \widetilde{\bf D}^{-1} \left(
\begin{array}{c}
\hat{d}_{N-1\,+}(\omega)
\\[.5ex]
\hat{d}_{N-1\,-}(\omega)
\end{array}
\right)\!,
\label{3.74}
\end{eqnarray}
from which we deduce that
\begin{eqnarray}
\lefteqn{
\left(
\begin{array}{c}
\hat{a}_{1-}(x_{1},\omega)
\\[.5ex]
\hat{a}_{N+}(x_{N-1},\omega)
\end{array}
\right) = \widetilde{\bf T}^{(N-2)} \left(
\begin{array}{c}
\hat{a}_{1+}(x_1,\omega)
\\[.5ex]
\hat{a}_{N-}(x_{N-1},\omega)
\end{array}
\right)
}
\nonumber
\\[.5ex]
& &
\hspace{3ex}
+ \, \widetilde{\bf A}' \left(
\begin{array}{c}
\hat{g}^{(N-3)}_{+}(\omega)
\\[.5ex]
\hat{g}^{(N-3)}_{-}(\omega)
\end{array}
\right)
+ \widetilde{\bf A}'' \left(
\begin{array}{c}
\hat{d}_{N-1\,+}(\omega)
\\[.5ex]
\hat{d}_{N-1\,-}(\omega)
\end{array}
\right)\!,
\label{3.75}
\end{eqnarray}
where the matrices read as
\begin{equation}
\widetilde{\bf T}^{(N-2)}=
(SRP)_{21}^{-2}
\left(\!
\begin{array}{c@{\hspace{2ex}}c}
  -(SRP)_{22}  & 1
\\[.5ex]
  (SRP)_{12} (SRP)_{21} \! - \! (SRP)_{11} (SRP)_{22}  & (SRP)_{11}
\end{array}
\right)\!,
\label{3.76}
\end{equation}
\begin{eqnarray}
\lefteqn{
\widetilde{\bf A}'=
(SRP)_{21}^{-2}
}
\nonumber \\
&& \times
\left( \!
\begin{array}{cc}
-(SRQ)_{21} & - (SRQ)_{22}
\\[.5ex]
(SRQ)_{11} (SRP)_{21} \! - \! (SRQ)_{21} (SRP)_{11}
& (SRQ)_{12} (SRP)_{21} \! - \! (SRQ)_{22} (SRP)_{11}
\end{array}\!
\right) \! ,
\nonumber \\ &&
\label{3.77}
\end{eqnarray}
\begin{eqnarray}
\lefteqn{
\widetilde{\bf A}''  =
(SRP)_{21}^{-2}
}
\nonumber \\
&& \times
\left(\!
\begin{array}{cc}
 -(SD^{-1})_{21}  &  - (SD^{-1})_{22}
\\[.5ex]
 (SD^{-1})_{11} (SRP)_{21} \! - \! (SD^{-1})_{21} (SRP)_{11}
& (SD^{-1})_{12} (SRP)_{21} \! - \! (SD^{-1})_{22} (SRP)_{11}
\end{array}\!
\right) \!.
\nonumber \\ &&
\label{3.78}
\end{eqnarray}
Note that $ \hat{g}^{(N-3)}_{\pm}(\omega) $ and
$ \hat{d}_{N-1\,\pm}^\dagger (\omega) $ are commuting quantities,
because they refer to different space intervals [cf. the commutation
relations (\ref{2.16}) or (\ref{2.25})].

Finally, we introduce bosonic operators $ \hat{g}^{(N-2)}_{\pm}(\omega) $
as linear combinations of the operators $ \hat{g}^{(N-3)}_{\pm}(\omega) $
and $\hat{d}_{N-1\,\pm}(\omega) $:
\begin{equation}
\left(
\begin{array}{c}
\hat{g}^{(N-2)}_{+}(\omega)
\\[.5ex]
\hat{g}^{(N-2)}_{-}(\omega)
\end{array}
\right) = \widetilde{\bf U} \left[ \widetilde{\bf A}' \left(
\begin{array}{c}
\hat{g}^{(N-3)}_{+}(\omega)
\\[.5ex]
\hat{g}^{(N-3)}_{-}(\omega)
\end{array}
\right)
+
\widetilde{\bf A}'' \left(
\begin{array}{c}
\hat{d}_{N-1\,+}(\omega)
\\[.5ex]
\hat{d}_{N-1\,-}(\omega)
\end{array}
\right)
\right]
\label{3.84}
\end{equation}
\begin{eqnarray}
\left[
\hat{g}^{(N-2)}_{\pm} (\omega),
    \big( \hat{g}^{(N-2)}_{\pm} (\omega') \big) ^{\dagger}
\right] & = & \delta( \omega - \omega' ) ,
\label{3.85}
\\
\left[
\hat{g}^{(N-2)}_{\pm} (\omega),
    \big( \hat{g}^{(N-2)}_{\mp} (\omega') \big) ^{\dagger}
\right] & = & 0.
\label{3.86}
\end{eqnarray}
The elements of the matrix $\widetilde{\bf U}$ are given by
\begin{eqnarray}
U_{11} & = &
\left\{
2 \left[
\mu_{+} + \left(
\frac{\mu_{+}} {\mu_{-}} \right)^{\!\frac{1}{2}}
\left| \mu_{0} \right|
\right]
\right\}^{-\frac{1}{2}}\hspace{-3ex},
\label{3.82}
\\
U_{21} & = &
-\left\{
2 \left[
\mu_{+} - \left(
\frac{\mu_{+}} {\mu_{-}} \right)^{\!\frac{1}{2}}
\left| \mu_{0} \right|
\right]
\right\}^{-\frac{1}{2}}\hspace{-3ex},
\label{3.82a}
\\
U_{12} & = &
 \exp{\left[i \arg{\!\left[
\mu_{0}
\right]}\right]} \left\{
2 \left[ \mu_{-} + \left(
\frac{\mu_{-}}{\mu_{+}} \right)^{\!\frac{1}{2}}
\left| \mu_{0} \right|
\right]
\right\}^{-\frac{1}{2}}\hspace{-3ex},
\label{3.83}
\\
U_{22} & = &
\exp{\left[i \arg{\!\left[
\mu_{0}
\right]}\right]} \left\{
2 \left[ \mu_{-} - \left(
\frac{\mu_{-}}{\mu_{+}} \right)^{\!\frac{1}{2}}
\left| \mu_{0} \right|
\right]
\right\}^{-\frac{1}{2}}\hspace{-3ex},
\label{3.83a}
\end{eqnarray}
where
\begin{eqnarray}
\mu_{+} & = & \left| A'_{11} \right|^2 + \left| A'_{12}
\right|^2
+ \left| A''_{11} \right|^2 + \left| A''_{12} \right|^2 \!,
\label{3.79}
\\
\mu_{-} & = & \left| A'_{21} \right|^2 + \left| A'_{22}
\right|^2
+ \left| A''_{21} \right|^2 + \left| A''_{22} \right|^2 \!,
\label{3.80}
\\
\mu_{0} & = & A'_{11} {A'}_{\! 21}^{*} + A'_{12}
{A'}_{\! 22}^{*}
+ A''_{11} {A''}_{\! 21}^{*}  + A''_{12} {A''}_{\!\! 22}^{*} .
\label{3.81}
\end{eqnarray}
Identifying the the inverse of $\widetilde{\bf U}$ with
$\widetilde{\bf A}^{(N-2)}$,
\begin{equation}
\widetilde{\bf A}^{(N-2)} = \widetilde{\bf U}^{-1}\hspace{-2ex},
\end{equation}
Eq.~(\ref{3.75}) [together with Eq.~(\ref{3.84})]
is the desired result (\ref{3.27}). From the structure of the
Green function [cf. Eqs.~(\ref{3.2}) -- (\ref{3.4a})]
it is clear that the input operators $\hat{a}_{1+}(x,\omega)$
and $\hat{a}_{N-}(x,\omega)$, respectively, can always be written in
the form given in Eqs.~(\ref{3.10a}) and (\ref{3.10b}) [with arbitrary $N$
($N\!\geq\!3$) in place of $N\!=\!3$]. With regard to the
basic-field operators $\hat{f}(x,\omega)$, both the input operators
and the noise operators $\hat{g}_\pm^{(N-2)}(\omega)$ refer
to different space intervals. Hence, all the commutation relations
(\ref{3.27i}) -- (\ref{3.27k}) are satisfied for arbitrary $N$
($N\!\geq\!3$), which implies that the commutation relations (\ref{3.17})
-- (\ref{3.18}) remain also valid when $\hat{a}_{N+}(x,\omega),
\hat{a}_{N+}^\dagger(x',\omega)$ and $\widetilde{\bf T}^{(N-2)},
\widetilde{\bf A}^{(N-2)}$ are substituted for
$\hat{a}_{3+}(x,\omega),\hat{a}_{3+}^\dagger(x',\omega)$
and $\widetilde{\bf T}^{(1)}, \widetilde{\bf A}^{(1)}$, respectively.
We finally mention that in a way similar to that outlined above for
the input--output relations the extension of the relations (\ref{3.18a})
and (\ref{3.18b}) to arbitrary $N$ ($N\!\geq\!3$) may be proved correct.

\section{Output correlation functions}
\label{illum}
\setcounter{equation}{0}
When the dielectric plate is in thermal equilibrium the density
operator of the matter excitations may be given by
\begin{equation}
\hat{\varrho}_{\rm dp} =
\exp\!\left\{
- \sum_{i=1}^2 \int_0^{\infty} {\rm d}\omega \,
\ln [ 1 + n_{\rm th}(\omega) ] \,
\frac{\hbar \omega}{k_{\rm B} T}\,
\hat{g}_i^{\dagger} (\omega) \hat{g}_i (\omega)
\right\}\!,
\label{D1}
\end{equation}
with $n_{\rm th}(\omega)$ from Eq.~(\ref{3.32h3a}).
Note that this density operator of course corresponds to the
thermal-equilibrium state of the basic field $\hat{f}(x,\omega)$
inside the plate.
The correlation functions (\ref{3.32d}) are then calculated to be
\begin{equation}
{\underline{\Gamma}}_{\{i_\mu\}}^{(p,q)} \!
\left( \big\{ \omega_\mu \big\} \right)
= {\delta}_{pq} \prod\limits_{\zeta=1}^p  \delta_{i_{\zeta} i_{\zeta+p}}
n_{\rm th}(\omega_{\zeta})\,
\delta( \omega_{\zeta} - \omega_{\zeta+p}),
\label{D2}
\end{equation}
where the set of indices $ \zeta $ denotes a permutation of the set of
indices $ \mu $,
so that $ \omega_{\zeta - 1} < \omega_{\zeta} $ for $ 2 \le \zeta \le p $
or $ p+2 \le \zeta \le p+q $ (note that
$ {\underline{\Gamma}}_{\{i_\mu\}}^{(p,q)} \!
\left( \big\{ \omega_\mu \big\} \right) $ vanishes if for any $ \zeta $ with
$ p \ge \zeta \ge 2 $ the
relation $ \omega_{\zeta-1} = \omega_{\zeta} $ is fulfilled).

We now appropriately label the terms that (after disentangling) occur on the
right-hand side of Eq.~(\ref{3.32b}).
For this purpose we introduce the set $ S^{(m,n)} $ of arrangements
of the $ m+n $ indices $ \zeta $  (disposed in ascending order) by
assigning them to two classes ($K=1,2$) and four (possibly empty)
groups ($j\!=\!1,2,3,4$). The class indices $K\!=\!1$ and $K\!=\!2$ refer
to the creation and destruction operators, respectively, and the group
indices $j\!=\!1,2,3,4$ are used to distinguish between the four excitations
to be considered. For $K\!=\!1\,(2)$ the indices refer for $j\!=\!1$ to
$ \hat{a}^{\dagger}_1 $ ($ \hat{a}_1 $), for $j\!=\!2$ to
$ \hat{a}^{\dagger}_2 $ ($ \hat{a}_2 $), for $j\!=\!3$ to
$ \hat{g}^{\dagger}_1 $ ($ \hat{g}_1 $), and for $j\!=\!4$ to
$ \hat{g}^{\dagger}_2 $ ($ \hat{g}_2 $).
The $ \lambda_j^K $ indices of the $K$th class and $j$th group
are denoted by $ \zeta_j^K\!(i) $, $i\!=\!1,\ldots,\lambda_j^K$.
 From Eq.~(\ref{3.32b}) we then find that
\begin{eqnarray}
\lefteqn{
{\underline{C}'}_{\{i_\mu\}}^{(m,n)}
\big(\big\{ \omega_\mu \big\}\big)
=\sum_{S^{(m,n)} }
{\underline{C}}_{\{i_{\zeta_1}\!;i_{\zeta_2}\}}
^{(\lambda_1^1,\lambda_1^2;\lambda_2^1,\lambda_2^2)}
\big(\big\{\omega_{\zeta_1};\omega_{\zeta_2}\big\}\big)
{\underline{\Gamma}}_{\{i_{\zeta_3}\}}^{(\lambda_3^1,\lambda_3^2)} \!
\left( \big\{ \omega_{\zeta_3} \big\} \right)
{\underline{\Gamma}}_{\{i_{\zeta_4}\}}^{(\lambda_4^1,\lambda_4^2)} \!
\left( \big\{ \omega_{\zeta_4} \big\} \right)
}
\nonumber
\\ & &
\times \,
\prod\limits_{\alpha=1}^{\lambda_1^1}
\prod\limits_{\beta=1}^{\lambda_2^1}
\prod\limits_{\gamma=1}^{\lambda_1^2}
\prod\limits_{\eta=1}^{\lambda_2^2}
T^*_{i_{\zeta^1_1 (\alpha)} 1} \big( \omega_{\zeta_{1}^1 (\alpha)} \big) \,
T^*_{i_{\zeta_2^1 (\beta)}  2} \big( \omega_{\zeta_{2}^1 (\beta)} \big) \,
T_{i_{\zeta_1^2 (\gamma)} 1} \big( \omega_{\zeta_{1}^2 (\gamma)} \big) \,
T_{i_{\zeta_2^2 (\eta)} 2} \big( \omega_{\zeta_{2}^2 (\eta)} \big)
\nonumber
\\ & &
\times
\prod\limits_{\phi=1}^{\lambda_3^1}
\left| A_{i_{\zeta_{3}^1(\phi)} 1}
\big(\omega_{\zeta_{3}^1(\phi)}\big) \right|^2
\prod\limits_{\chi=1}^{\lambda_4^1}
\left| A_{i_{\zeta_{4}^1(\chi)} 2}
\big(\omega_{\zeta_{4}^1(\chi)}\big) \right|^2\!,
\label{D4}
\end{eqnarray}
where
\begin{eqnarray}
\lefteqn{
{\underline{C}}_{\{i_{\zeta_1}\!;i_{\zeta_2}\}}
^{(\lambda_1^1,\lambda_1^2;\lambda_2^1,\lambda_2^2)} \!
\big(\big\{\omega_{\zeta_1};\omega_{\zeta_2}\big\}\big)
}
\nonumber \\
&& \hspace{2ex}
=
\left\langle
\left[
\prod\limits_{\alpha=1}^{\lambda_1^1}
\hat{a}_1^{\dagger} \big( \omega_{\zeta_1^1 (\alpha)} \big)
\right]
\left[
\prod\limits_{\beta=1}^{\lambda_2^1}
\hat{a}_2^{\dagger} \big( \omega_{\zeta_2^1 (\beta)} \big)
\right]
\left[
\prod\limits_{\gamma=1}^{\lambda_1^2}
\hat{a}_1 \big( \omega_{\zeta_1^2 (\gamma)} \big)
\right]
\left[
\prod\limits_{\eta=1}^{\lambda_2^2}
\hat{a}_2 \big( \omega_{\zeta_2^2 (\eta)} \big)
\right]
\right\rangle
\hspace{5ex}
\label{D5}
\end{eqnarray}
(the notation $\zeta_j$ is used to indicate sets of indices,
without distinguishing between the two classes).
\end{appendix}

\newpage

\vspace*{\fill}

\unitlength1mm
\begin{picture}(150,70)
\put(-10,-201){\psfig{figure=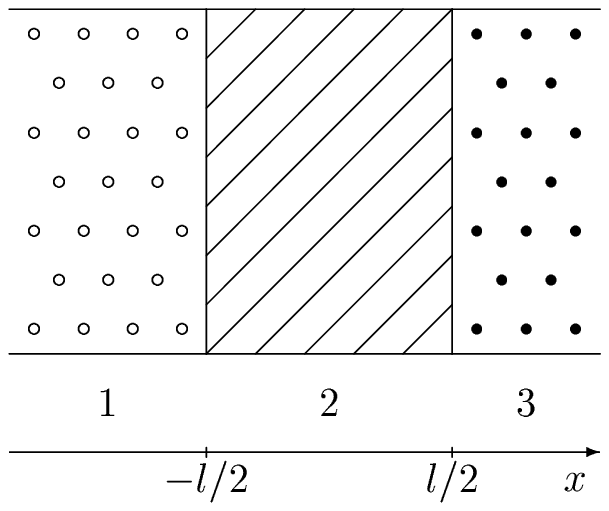,height=12.5in,angle=0}}
\end{picture}
\\
\noindent
{\bf Figure 1}:
Scheme of the single-slab dielectric plate (2) of thickness $l$
embedded in dielectric matter  (1 and 3).
\unitlength1mm

\vspace*{\fill}

\newpage

\vspace*{\fill}

\begin{picture}(150,150)
\put(-55,-140){\psfig{figure=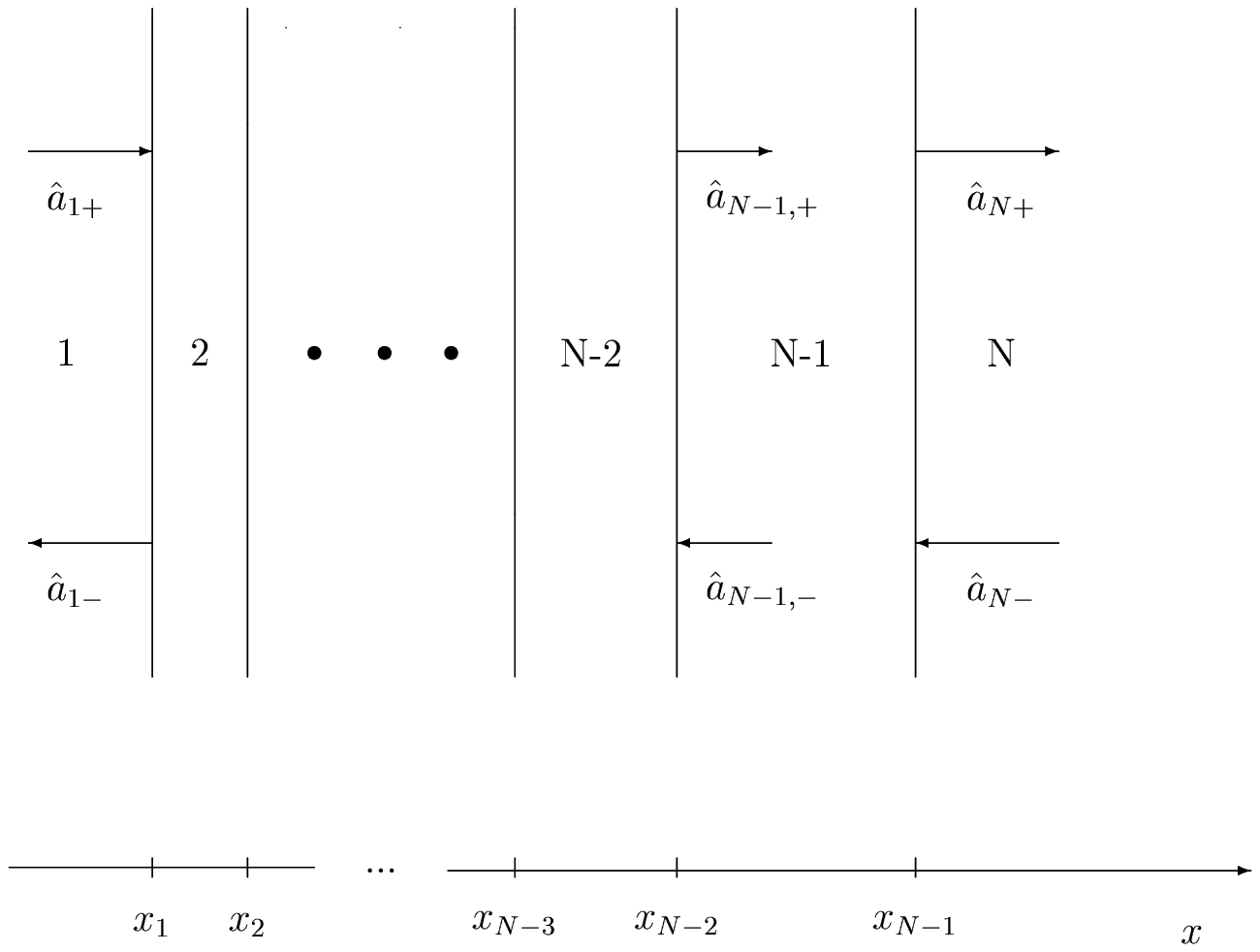,height=12.5in,angle=0}}
\end{picture}
\\
\noindent
{\bf Figure 2}:
Scheme of the multilayer dielectric configuration, the arrows
together with the amplitude operators indicating incoming
and outgoing fields.

\vspace*{\fill}

\newpage

\vspace*{\fill}

\begin{picture}(150,150)
\put(-35,-100){\psfig{figure=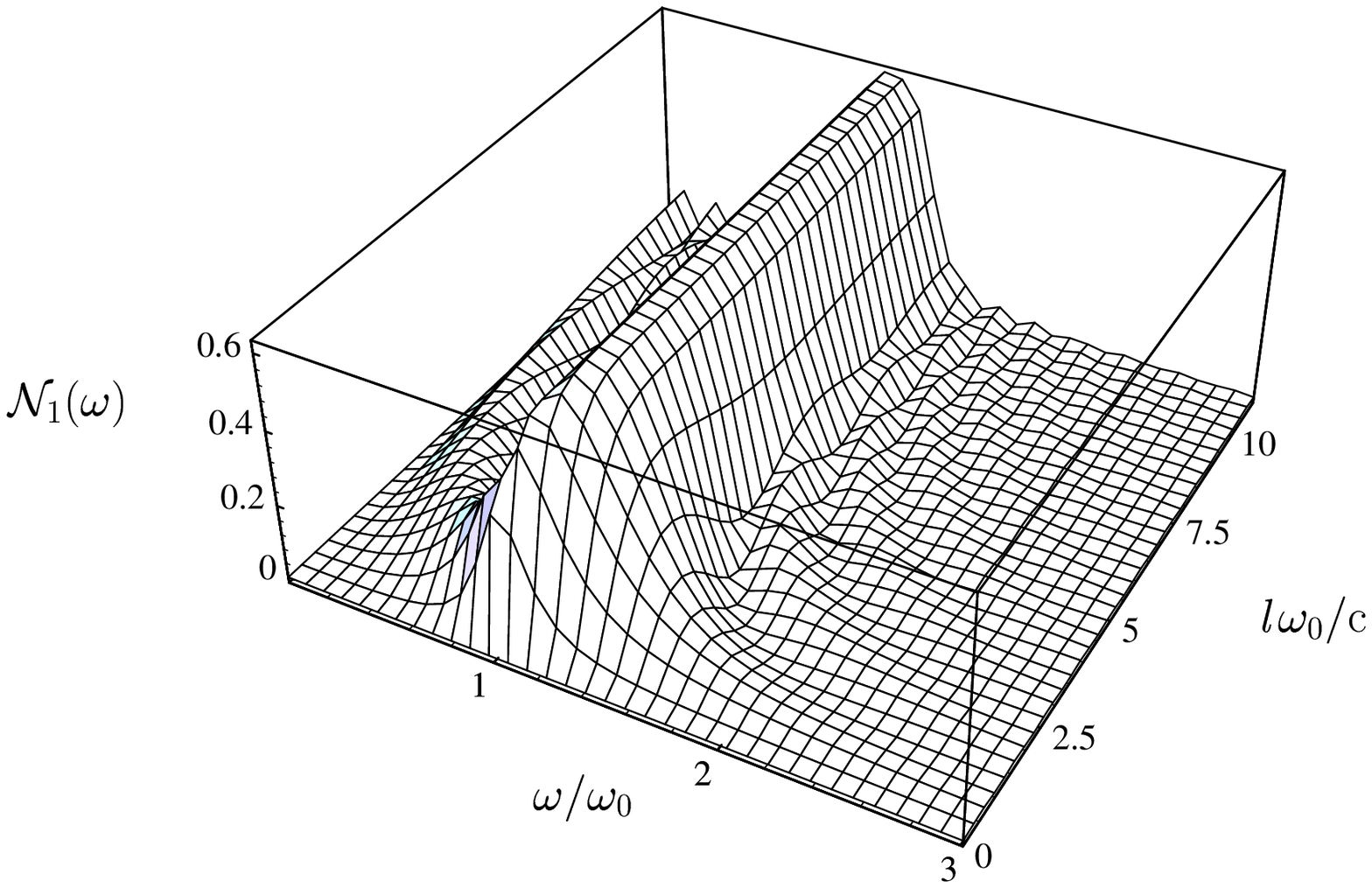,height=12.5in,angle=0}}
\end{picture}
\\
\noindent
{\bf Figure 3}:
The ratio of photon-number densities of the reflected outgoing field
and the incoming field, ${\cal N}_{1}(\omega)
 =  N'_{{\rm ph} \, 1}(\omega)/N_{{\rm ph} \, 1}(\omega)$,
as a function of frequency and plate thickness
for a single-resonance medium ($ \omega_0  =  \omega_1 $,
$ \Gamma  =  0.1 \omega_0 $).

\vspace*{\fill}

\newpage

\vspace*{\fill}

\begin{picture}(150,150)
\put(-35,-100){\psfig{figure=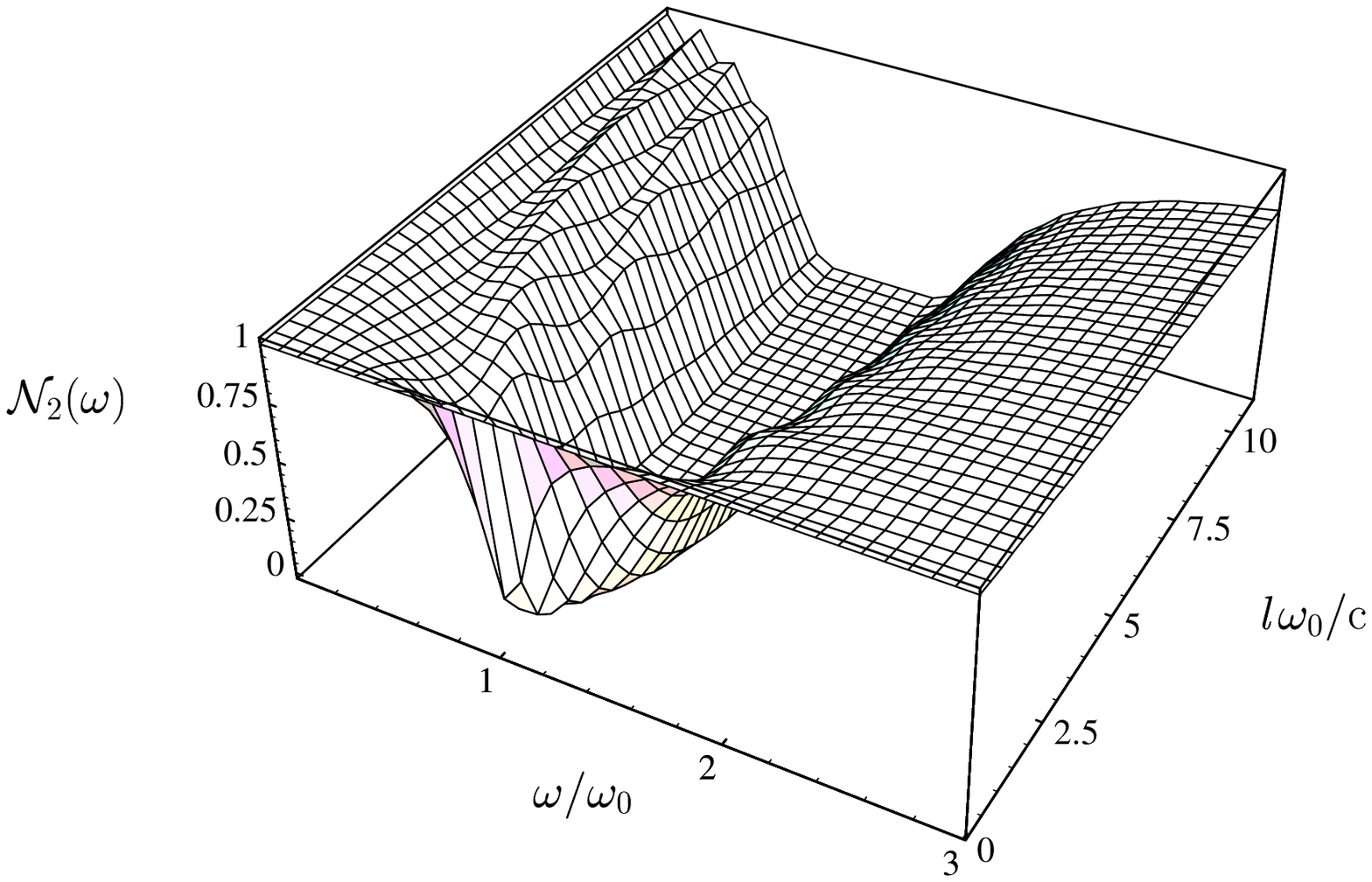,height=12.5in,angle=0}}
\end{picture}
\\
\noindent
{\bf Figure 4}:
The ratio of the photon-number densities of the transmitted outgoing field
and the incoming field, $ {\cal N}_{2} (\omega)
 = N'_{{\rm ph} \, 2}(\omega) /N_{{\rm ph} \, 1} (\omega) $,
as a function of frequency and plate thickness
for a single-resonance medium ($ \omega_0 = \omega_1 $,
$ \Gamma = 0.1 \omega_0 $).

\vspace*{\fill}

\newpage

\vspace*{\fill}

\begin{picture}(150,150)
\put(-35,-100){\psfig{figure=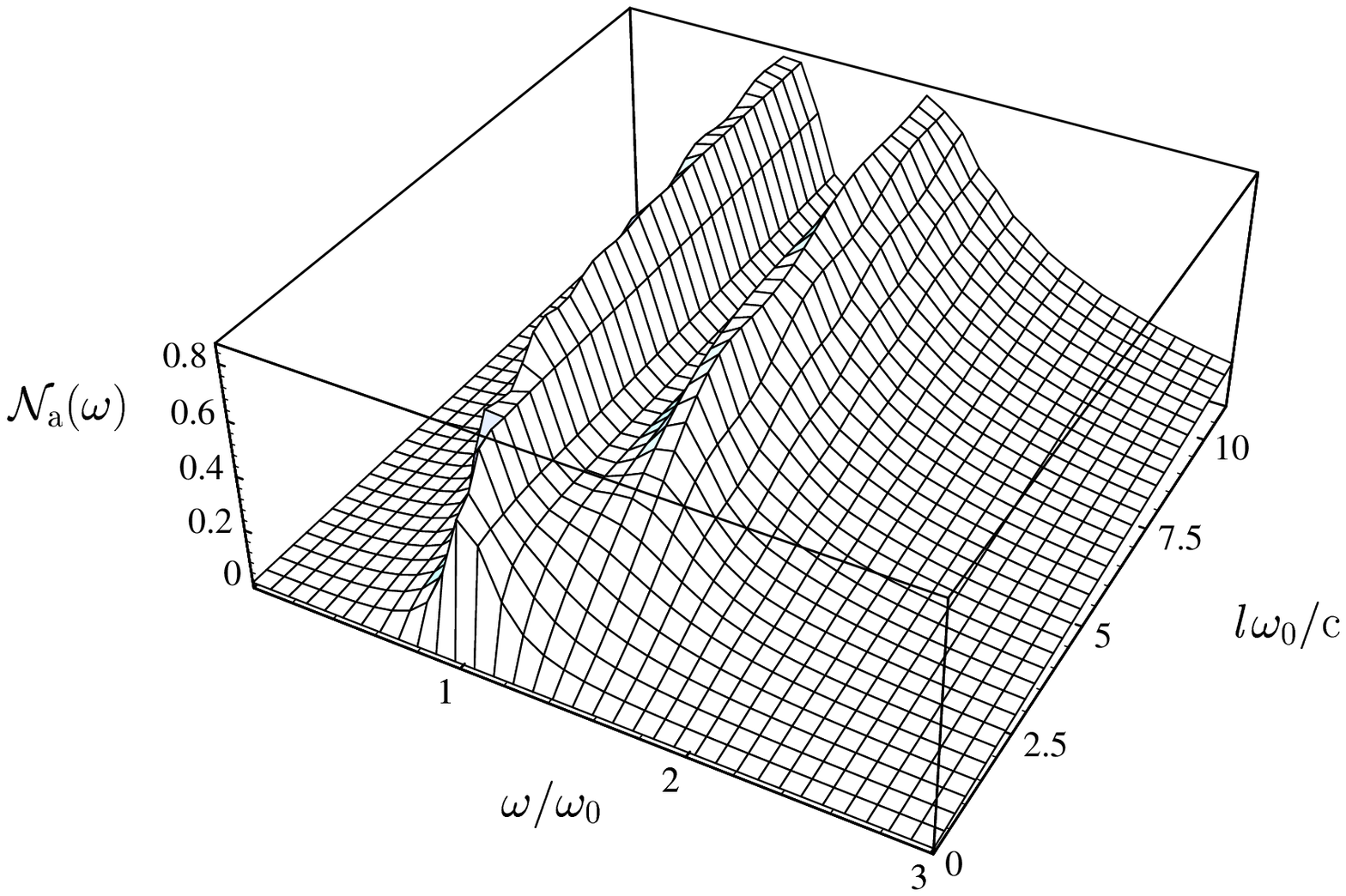,height=12.5in,angle=0}}
\end{picture}
\\
\noindent
{\bf Figure 5}:
The ratio of the photon-number densities of the radiation
absorbed in the plate and the incoming field, ${\cal N}_{\rm a}
 = [N_{{\rm ph}\,1}(\omega) - (N'_{{\rm ph}\,1}(\omega) +
N'_{{\rm ph}\,2}(\omega))] /N_{{\rm ph}\,1}(\omega) =
\alpha_1(\omega) = \alpha_2(\omega)$, as a function of frequency
and plate thickness for a single-resonance medium ($ \omega_0 = \omega_1 $,
$ \Gamma = 0.1 \omega_0 $).

\vspace*{\fill}

\newpage

\vspace*{\fill}

\begin{picture}(150,150)(17,0)
\put(-35,-100){\psfig{figure=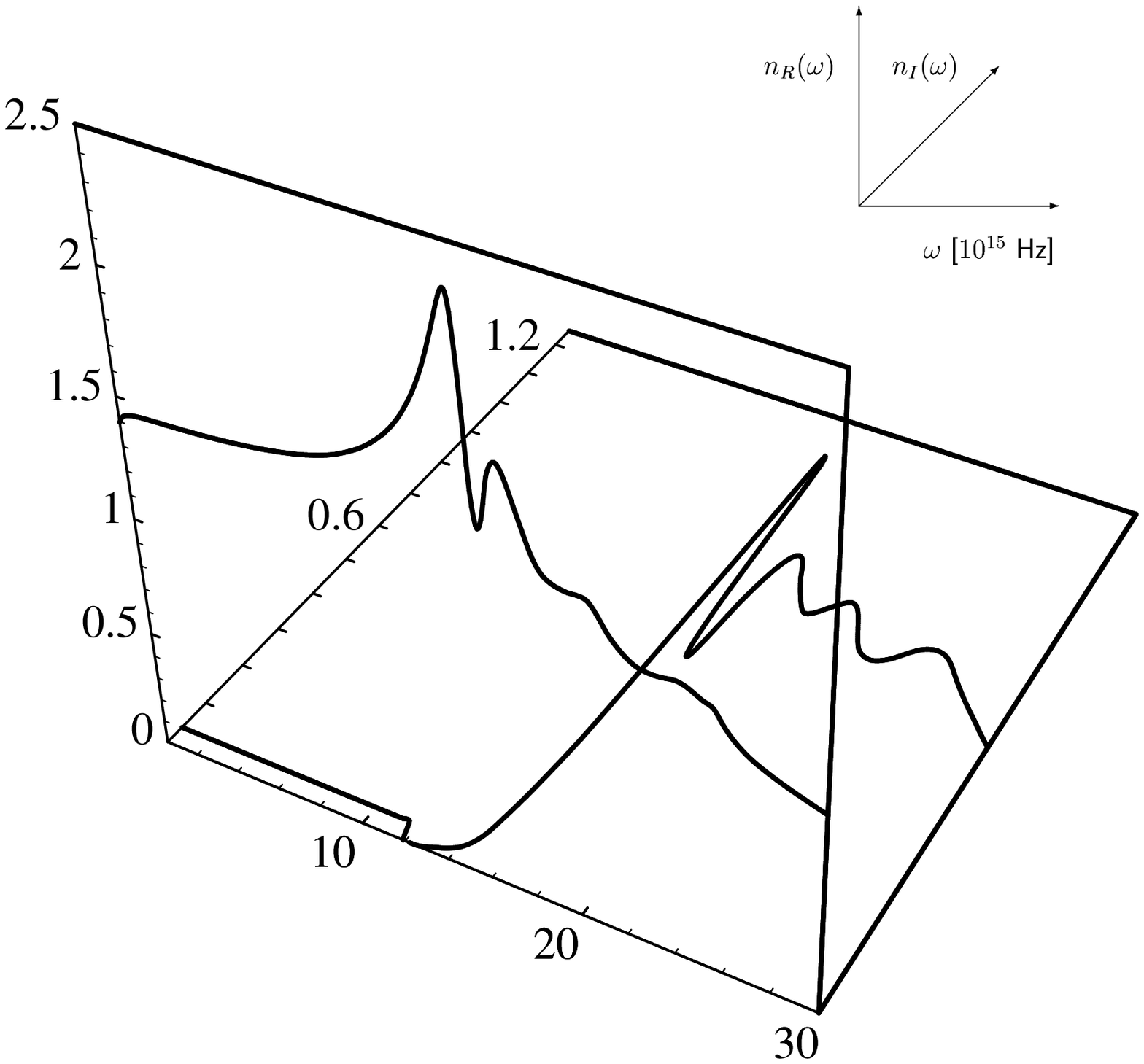,height=12.5in,angle=0}}
\end{picture}
\vspace{2.5cm}
\\
\noindent
{\bf Figure 6}:
Real and imaginary parts of the refractive index
of amorphic SiO$_2$ as a function of frequency.

\vspace*{\fill}

\newpage

\vspace*{\fill}

\begin{picture}(150,150)
\put(-35,-100){\psfig{figure=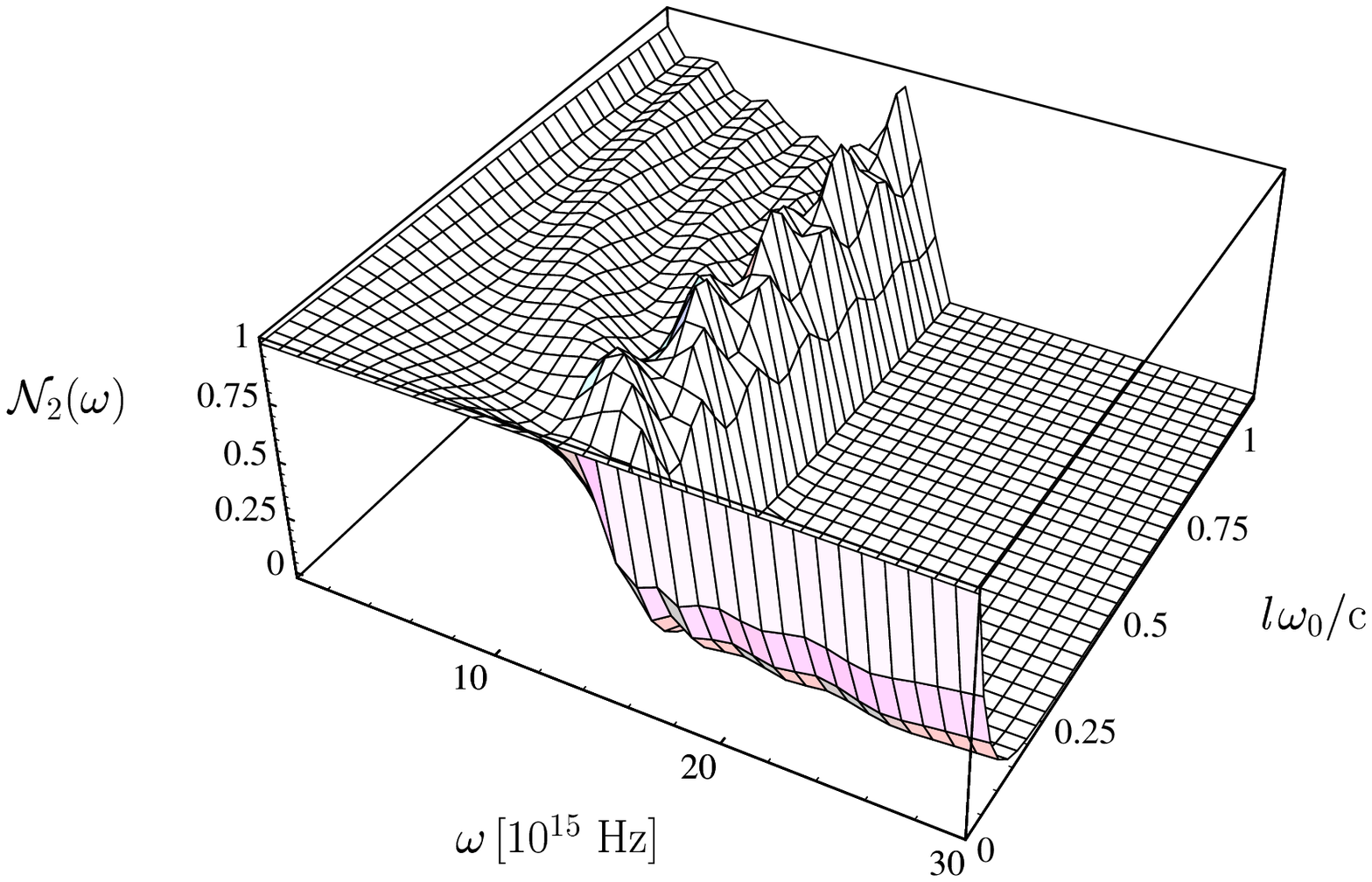,height=12.5in,angle=0}}
\end{picture}
\\
\noindent
{\bf Figure 7}:
The ratio of the photon-number densities of the transmitted outgoing field
and the incoming field,  $ {\cal N}_{2} (\omega)
 = N'_{{\rm ph} \, 2}(\omega) /N_{{\rm ph} \, 1} (\omega) $,
as a function of frequency and plate thickness
for amorphic SiO$_2$ ($\omega_0 = 1 \cdot 10^{15} $ Hz).

\vspace*{\fill}

\end{document}